\documentclass[lettersize,journal]{IEEEtran}
\usepackage{array}
\usepackage{stfloats}
\usepackage{enumitem}
\usepackage{url}
\usepackage{verbatim}
\usepackage{amsmath,amssymb,amsfonts}
\usepackage{graphicx}
\usepackage{subcaption} 
\usepackage{textcomp}
\def\BibTeX{{\rm B\kern-.05em{\sc i\kern-.025em b}\kern-.08em
    T\kern-.1667em\lower.7ex\hbox{E}\kern-.125emX}}
\usepackage{cite}
\usepackage{lineno}
\usepackage{booktabs}
\usepackage{balance} 
\usepackage{tabularx}
\usepackage{mathtools}
\usepackage{graphics}
\usepackage{epstopdf}
\usepackage{amsmath,amssymb,amsfonts}
\usepackage[english]{babel}
\usepackage{amsthm}
\usepackage{soul}
\usepackage{color}
\usepackage{caption}
\usepackage{wrapfig}
\usepackage{comment}
\usepackage{float}
\usepackage{amssymb}
\usepackage{lipsum}
\usepackage{enumitem}
\usepackage{algorithm}
\usepackage{algorithmic}
\usepackage[acronyms,nonumberlist,nopostdot,nomain,nogroupskip,acronymlists={hidden}]{glossaries}
\usepackage{enumerate}   
\usepackage[inkscapelatex=false]{svg}
\usepackage[english]{babel}
\usepackage[acronyms]{glossaries}
\newacronym{collprob}{CP}{Collision Probability}
\newacronym{pcr}{PCR}{Packet Collision Ratio}
\newacronym{cbr}{CBR}{Channel Bussy Ration}
\newacronym{pir}{PIR}{Packet Inter-Reception}
\newacronym{ud}{UD}{Update Delay}
\newacronym{sr}{SR}{Scheduling Request}
\newacronym{dg}{DG}{Distrubuted Grant}
\newacronym{prr}{PRR}{Packet Reception Ratio}
\newacronym{ma}{MA}{Multi-Agent}
\newacronym{dqn}{DQN}{Deep Q-Network}
\newacronym{madrl}{MADRL}{Multi-Agent Deep Reinforcement Learning}
\newacronym{marl}{MARL}{Multi-Agent Reinforcement Learning}
\newacronym{psbch}{PSBCH}{Physical Sidelink Broadcast Channel}
\newacronym{psfch}{PSFCH}{Physical Sidelink Feedback Channel}
\newacronym{pscch}{PSCCH}{Physical Sidelink Control Channel}
\newacronym{pssch}{PSSCH}{Physical Sidelink Shared Channel}
\newacronym{scs}{SCS}{Subcarrier Spacing}
\newacronym{ndi}{NDI}{New Data Indicator}
\newacronym{rc}{RC}{re-selection counter}
\newacronym{rp}{RP}{Resource Pool}
\newacronym{rri}{RRI}{Resource Reservation Interval}
\newacronym{rsrp}{RSRP}{Reference Signal Received Power}

\newacronym{rnn}{RNN}{Recurrent Neural Networks}
\newacronym{ra}{RA}{Resource Allocation}
\newacronym{dra}{DRA}{Distributed Resource Allocation}
\newacronym{rsvp}{P$_{rsvp}$}{Resource Reservation Period}
\newacronym{sci}{SCI}{Sidelink Control Information}
\newacronym{rl}{RL}{Reinforcement Learning}
\newacronym{ppbp}{PPBP}{Poisson Pareto Burst Process}
\newacronym{pdr}{PDR}{Packet Delivery Ratio}

\newacronym{hu}{HU}{Heavy User}
\newacronym{dnn}{DNN}{Deep-Neural Network}
\newacronym{relu}{ReLU}{Rectified Linear Unit}
\newacronym{ns3}{ns-3}{Network Simulator 3}
\newacronym{sl}{SL}{Sidelink}
\newacronym{rnti}{SL-RNTI}{Sidelink Radio Network Temporary Identifier}
\newacronym{3gpp}{3GPP}{3rd Generation Partnership Project}
\newacronym{4g}{4G}{4th generation}
\newacronym{5g}{5G}{5th generation}
\newacronym{6g}{6G}{6th generation}
\newacronym{5gc}{5GC}{5G Core}
\newacronym{isd}{ISD}{Intersite distance}

\newacronym{adc}{ADC}{Analog to Digital Converter}
\newacronym{aerpaw}{AERPAW}{Aerial Experimentation and Research Platform for Advanced Wireless}
\newacronym{ai}{AI}{Artificial Intelligence}
\newacronym{aimd}{AIMD}{Additive Increase Multiplicative Decrease}
\newacronym{am}{AM}{Acknowledged Mode}
\newacronym{amc}{AMC}{Adaptive Modulation and Coding}
\newacronym{amf}{AMF}{Access and Mobility Management Function}
\newacronym{aops}{AOPS}{Adaptive Order Prediction Scheduling}
\newacronym{api}{API}{Application Programming Interface}
\newacronym{apn}{APN}{Access Point Name}
\newacronym{ap}{AP}{Application Protocol}
\newacronym{aqm}{AQM}{Active Queue Management}
\newacronym{ausf}{AUSF}{Authentication Server Function}
\newacronym{avc}{AVC}{Advanced Video Coding}
\newacronym{awgn}{AGWN}{Additive White Gaussian Noise}
\newacronym{balia}{BALIA}{Balanced Link Adaptation Algorithm}
\newacronym{bbu}{BBU}{Base Band Unit}
\newacronym{bdp}{BDP}{Bandwidth-Delay Product}
\newacronym{ber}{BER}{Bit Error Rate}
\newacronym{bf}{BF}{Beamforming}
\newacronym{bler}{BLER}{Block Error Rate}
\newacronym{brr}{BRR}{Bayesian Ridge Regressor}
\newacronym{bs}{BS}{Base Station}
\newacronym{bsr}{BSR}{Buffer Status Report}
\newacronym{bss}{BSS}{Business Support System}
\newacronym{ca}{CA}{Carrier Aggregation}
\newacronym{caas}{CaaS}{Connectivity-as-a-Service}
\newacronym{cav}{CAV}{Connected and Autonomous Vehicle}
\newacronym{cb}{CB}{Code Block}
\newacronym{cc}{CC}{Congestion Control}
\newacronym{ccid}{CCID}{Congestion Control ID}
\newacronym{cco}{CC}{Carrier Component}
\newacronym{cd}{CD}{Continuous Delivery}
\newacronym{cdd}{CDD}{Cyclic Delay Diversity}
\newacronym{cdf}{CDF}{Cumulative Distribution Function}
\newacronym{cdn}{CDN}{Content Distribution Network}
\newacronym{cli}{CLI}{Command-line Interface}
\newacronym{cn}{CN}{Core Network}
\newacronym{codel}{CoDel}{Controlled Delay Management}
\newacronym{comac}{COMAC}{Converged Multi-Access and Core}
\newacronym{cord}{CORD}{Central Office Re-architected as a Datacenter}
\newacronym{cornet}{CORNET}{COgnitive Radio NETwork}
\newacronym{cosmos}{COSMOS}{Cloud Enhanced Open Software Defined Mobile Wireless Testbed for City-Scale Deployment}
\newacronym{cots}{COTS}{Commercial Off-the-Shelf}
\newacronym{cp}{CP}{Control Plane}
\newacronym{cyp}{CP}{Cyclic Prefix}
\newacronym{up}{UP}{User Plane}
\newacronym{cpu}{CPU}{Central Processing Unit}
\newacronym{cqi}{CQI}{Channel Quality Information}
\newacronym{cr}{CR}{Cognitive Radio}
\newacronym{cran}{CRAN}{Cloud \gls{ran}}
\newacronym{crs}{CRS}{Cell Reference Signal}
\newacronym{csi}{CSI}{Channel State Information}
\newacronym{csirs}{CSI-RS}{Channel State Information - Reference Signal}
\newacronym{cu}{CU}{Central Unit}
\newacronym{d2tcp}{D$^2$TCP}{Deadline-aware Data center TCP}
\newacronym{d3}{D$^3$}{Deadline-Driven Delivery}
\newacronym{dac}{DAC}{Digital to Analog Converter}
\newacronym{dag}{DAG}{Directed Acyclic Graph}
\newacronym{das}{DAS}{Distributed Antenna System}
\newacronym{dash}{DASH}{Dynamic Adaptive Streaming over HTTP}
\newacronym{dc}{DC}{Dual Connectivity}
\newacronym{dccp}{DCCP}{Datagram Congestion Control Protocol}
\newacronym{dce}{DCE}{Direct Code Execution}
\newacronym{dci}{DCI}{Downlink Control Information}
\newacronym{dctcp}{DCTCP}{Data Center TCP}
\newacronym{dl}{DL}{Downlink}
\newacronym{dmr}{DMR}{Deadline Miss Ratio}
\newacronym{dmrs}{DMRS}{DeModulation Reference Signal}
\newacronym{drlcc}{DRL-CC}{Deep Reinforcement Learning Congestion Control}
\newacronym{dsrc}{DSRC}
{Dedicated Short Range Communications}
\newacronym{d2d}{D2D}{device-to-device}
\newacronym{drs}{DRS}{Discovery Reference Signal}
\newacronym{du}{DU}{Distributed Unit}
\newacronym{e2e}{E2E}{end-to-end}
\newacronym{earfcn}{EARFCN}{E-UTRA Absolute Radio Frequency Channel Number}
\newacronym{ecaas}{ECaaS}{Edge-Cloud-as-a-Service}
\newacronym{ecn}{ECN}{Explicit Congestion Notification}
\newacronym{edf}{EDF}{Earliest Deadline First}
\newacronym{embb}{eMBB}{Enhanced Mobile Broadband}
\newacronym{empower}{EMPOWER}{EMpowering transatlantic PlatfOrms for advanced WirEless Research}
\newacronym{enb}{eNB}{evolved Node Base}
\newacronym{endc}{EN-DC}{E-UTRAN-\gls{nr} \gls{dc}}
\newacronym{epc}{EPC}{Evolved Packet Core}
\newacronym{eps}{EPS}{Evolved Packet System}
\newacronym{es}{ES}{Edge Server}
\newacronym{etsi}{ETSI}{European Telecommunications Standards Institute}
\newacronym[firstplural=Estimated Times of Arrival (ETAs)]{eta}{ETA}{Estimated Time of Arrival}
\newacronym{eutran}{E-UTRAN}{Evolved Universal Terrestrial Access Network}
\newacronym{faas}{FaaS}{Function-as-a-Service}
\newacronym{fapi}{FAPI}{Functional Application Platform Interface}
\newacronym{fdd}{FDD}{Frequency Division Duplexing}
\newacronym{fdm}{FDM}{Frequency Division Multiplexing}
\newacronym{fdma}{FDMA}{Frequency Division Multiple Access}
\newacronym{fed4fire}{FED4FIRE+}{Federation 4 Future Internet Research and Experimentation Plus}
\newacronym{fir}{FIR}{Finite Impulse Response}
\newacronym{fit}{FIT}{Future \acrlong{iot}}
\newacronym{fpga}{FPGA}{Field Programmable Gate Array}
\newacronym{fr2}{FR2}{Frequency Range 2}
\newacronym{fr1}{FR1}{Frequency Range 1}
\newacronym{fs}{FS}{Fast Switching}
\newacronym{fscc}{FSCC}{Flow Sharing Congestion Control}
\newacronym{ftp}{FTP}{File Transfer Protocol}
\newacronym{fw}{FW}{Flow Window}
\newacronym{ge}{GE}{Gaussian Elimination}
\newacronym{gnb}{gNB}{Next Generation Node Base}
\newacronym{gop}{GOP}{Group of Pictures}
\newacronym{gpr}{GPR}{Gaussian Process Regressor}
\newacronym{gps}{GPS}{Global Position System}
\newacronym{gpu}{GPU}{Graphics Processing Unit}
\newacronym{gtp}{GTP}{GPRS Tunneling Protocol}
\newacronym{gtpc}{GTP-C}{GPRS Tunnelling Protocol Control Plane}
\newacronym{gtpu}{GTP-U}{GPRS Tunnelling Protocol User Plane}
\newacronym{gtpv2c}{GTPv2-C}{\gls{gtp} v2 - Control}
\newacronym{gw}{GW}{Gateway}
\newacronym{harq}{HARQ}{Hybrid Automatic Repeat reQuest}
\newacronym{hetnet}{HetNet}{Heterogeneous Network}
\newacronym{hh}{HH}{Hard Handover}
\newacronym{hol}{HOL}{Head-of-Line}
\newacronym{hqf}{HQF}{Highest-quality-first}
\newacronym{hss}{HSS}{Home Subscription Server}
\newacronym{http}{HTTP}{HyperText Transfer Protocol}
\newacronym{ia}{IA}{Initial Access}
\newacronym{iab}{IAB}{Integrated Access and Backhaul}
\newacronym{ic}{IC}{Incident Command}
\newacronym{ietf}{IETF}{Internet Engineering Task Force}
\newacronym{imsi}{IMSI}{International Mobile Subscriber Identity}
\newacronym{imt}{IMT}{International Mobile Telecommunication}
\newacronym{iot}{IoT}{Internet of Things}
\newacronym{ip}{IP}{Internet Protocol}
\newacronym{itu}{ITU}{International Telecommunication Union}
\newacronym{kpi}{KPI}{Key Performance Indicator}
\newacronym{kpm}{KPM}{Key Performance Measurement}
\newacronym{kvm}{KVM}{Kernel-based Virtual Machine}
\newacronym{los}{LoS}{Line of Sight}
\newacronym{lsm}{LSM}{Link-to-System Mapping}
\newacronym{lstm}{LSTM}{Long Short Term Memory}
\newacronym{lte}{LTE}{Long Term Evolution}
\newacronym{lxc}{LXC}{Linux Container}
\newacronym{m2m}{M2M}{Machine to Machine}
\newacronym{mac}{MAC}{Medium Access Control}
\newacronym{manet}{MANET}{Mobile Ad Hoc Network}
\newacronym{mano}{MANO}{Management and Orchestration}
\newacronym{mc}{MC}{Multi-Connectivity}
\newacronym{mcc}{MCC}{Mobile Cloud Computing}
\newacronym{mchem}{MCHEM}{Massive Channel Emulator}
\newacronym{mcs}{MCS}{Modulation and Coding Scheme}
\newacronym{mec2}{MEC}{Multi-access Edge Computing}
\newacronym{mec}{MEC}{Mobile Edge Computing}
\newacronym{mfc}{MFC}{Mobile Fog Computing}
\newacronym{mgen}{MGEN}{Multi-Generator}
\newacronym{mi}{MI}{Mutual Information}
\newacronym{mib}{MIB}{Master Information Block}
\newacronym{miesm}{MIESM}{Mutual Information Based Effective SINR}
\newacronym{mimo}{MIMO}{Multiple Input, Multiple Output}
\newacronym{ml}{ML}{Machine Learning}
\newacronym{mlr}{MLR}{Maximum-local-rate}
\newacronym[plural=\gls{mme}s,firstplural=Mobility Management Entities (MMEs)]{mme}{MME}{Mobility Management Entity}
\newacronym{mmtc}{mMTC}{Massive Machine-Type Communications}
\newacronym{mmwave}{mmWave}{millimeter wave}
\newacronym{mpdccp}{MP-DCCP}{Multipath Datagram Congestion Control Protocol}
\newacronym{mptcp}{MPTCP}{Multipath TCP}
\newacronym{mr}{MR}{Maximum Rate}
\newacronym{mrdc}{MR-DC}{Multi \gls{rat} \gls{dc}}
\newacronym{mse}{MSE}{Mean Square Error}
\newacronym{mss}{MSS}{Maximum Segment Size}
\newacronym{mt}{MT}{Mobile Termination}
\newacronym{mtd}{MTD}{Machine-Type Device}
\newacronym{mtu}{MTU}{Maximum Transmission Unit}
\newacronym{mumimo}{MU-MIMO}{Multi-user \gls{mimo}}
\newacronym{mvno}{MVNO}{Mobile Virtual Network Operator}
\newacronym{nalu}{NALU}{Network Abstraction Layer Unit}
\newacronym{nas}{NAS}{Network Attached Storage}
\newacronym{nat}{NAT}{Network Address Translation}
\newacronym{nbiot}{NB-IoT}{Narrow Band IoT}
\newacronym{nfv}{NFV}{Network Function Virtualization}
\newacronym{nfvi}{NFVI}{Network Function Virtualization Infrastructure}
\newacronym{ni}{NI}{Network Interfaces}
\newacronym{nic}{NIC}{Network Interface Card}
\newacronym{now}{NOW}{Non Overlapping Window}
\newacronym{nsm}{NSM}{Network Service Mesh}
\newacronym{nr}{NR}{New Radio}
\newacronym{nrf}{NRF}{Network Repository Function}
\newacronym{nsa}{NSA}{Non Stand Alone}
\newacronym{nse}{NSE}{Network Slicing Engine}
\newacronym{nssf}{NSSF}{Network Slice Selection Function}
\newacronym{o2i}{O2I}{Outdoor to Indoor}
\newacronym{oai}{OAI}{OpenAirInterface}
\newacronym{oaicn}{OAI-CN}{\gls{oai} \acrlong{cn}}
\newacronym{oairan}{OAI-RAN}{\acrlong{oai} \acrlong{ran}}
\newacronym{oam}{OAM}{Operations, Administration and Maintenance}
\newacronym{ofdm}{OFDM}{Orthogonal Frequency Division Multiplexing}
\newacronym{olia}{OLIA}{Opportunistic Linked Increase Algorithm}
\newacronym{omec}{OMEC}{Open Mobile Evolved Core}
\newacronym{onap}{ONAP}{Open Network Automation Platform}
\newacronym{onf}{ONF}{Open Networking Foundation}
\newacronym{onos}{ONOS}{Open Networking Operating System}
\newacronym{oom}{OOM}{\gls{onap} Operations Manager}
\newacronym{opnfv}{OPNFV}{Open Platform for \gls{nfv}}
\newacronym{oran}{O-RAN}{Open RAN}
\newacronym{orbit}{ORBIT}{Open-Access Research Testbed for Next-Generation Wireless Networks}
\newacronym{os}{OS}{Operating System}
\newacronym{oss}{OSS}{Operations Support System}
\newacronym{pa}{PA}{Position-aware}
\newacronym{pase}{PASE}{Prioritization, Arbitration, and Self-adjusting Endpoints}
\newacronym{pawr}{PAWR}{Platforms for Advanced Wireless Research}
\newacronym{pbch}{PBCH}{Physical Broadcast Channel}
\newacronym{pcef}{PCEF}{Policy and Charging Enforcement Function}
\newacronym{pcfich}{PCFICH}{Physical Control Format Indicator Channel}
\newacronym{pcrf}{PCRF}{Policy and Charging Rules Function}
\newacronym{pdcch}{PDCCH}{Physical Downlink Control Channel}
\newacronym{pdcp}{PDCP}{Packet Data Convergence Protocol}
\newacronym{pdsch}{PDSCH}{Physical Downlink Shared Channel}
\newacronym{pdu}{PDU}{Packet Data Unit}
\newacronym{pf}{PF}{Proportional Fair}
\newacronym{pgw}{PGW}{Packet Gateway}
\newacronym{phich}{PHICH}{Physical Hybrid ARQ Indicator Channel}
\newacronym{phy}{PHY}{Physical}
\newacronym{pmch}{PMCH}{Physical Multicast Channel}
\newacronym{pmi}{PMI}{Precoding Matrix Indicators}
\newacronym{powder}{POWDER}{Platform for Open Wireless Data-driven Experimental Research}
\newacronym{ppo}{PPO}{Proximal Policy Optimization}
\newacronym{ppp}{PPP}{Poisson Point Process}
\newacronym{prach}{PRACH}{Physical Random Access Channel}
\newacronym{rb}{RB}{Physical Resource Block}
\newacronym{prb}{PRB}{Physical Resource Block}
\newacronym{psnr}{PSNR}{Peak Signal to Noise Ratio}
\newacronym{pss}{PSS}{Primary Synchronization Signal}
\newacronym{pucch}{PUCCH}{Physical Uplink Control Channel}
\newacronym{pusch}{PUSCH}{Physical Uplink Shared Channel}
\newacronym{qam}{QAM}{Quadrature Amplitude Modulation}
\newacronym{qci}{QCI}{\gls{qos} Class Identifier}
\newacronym{qoe}{QoE}{Quality of Experience}
\newacronym{qos}{QoS}{Quality of Service}
\newacronym{quic}{QUIC}{Quick UDP Internet Connections}
\newacronym{rach}{RACH}{Random Access Channel}
\newacronym{ran}{RAN}{Radio Access Network}
\newacronym{rbg}{RBG}{Resource Block Group}
\newacronym{rcn}{RCN}{Research Coordination Network}
\newacronym{rec}{REC}{Radio Edge Cloud}
\newacronym{red}{RED}{Random Early Detection}
\newacronym{renew}{RENEW}{Reconfigurable Eco-system for Next-generation End-to-end Wireless}
\newacronym{rf}{RF}{Radio Frequency}
\newacronym{rfc}{RFC}{Request for Comments}
\newacronym{rfr}{RFR}{Random Forest Regressor}
\newacronym{ric}{RIC}{RAN Intelligent Controller}
\newacronym{rlc}{RLC}{Radio Link Control}
\newacronym{rlf}{RLF}{Radio Link Failure}
\newacronym{rlnc}{RLNC}{Random Linear Network Coding}
\newacronym{rmr}{RMR}{RIC Message Router}
\newacronym{rmse}{RMSE}{Root Mean Squared Error}
\newacronym{rnis}{RNIS}{Radio Network Information Service}
\newacronym{rr}{RR}{Round Robin}
\newacronym{rrc}{RRC}{Radio Resource Control}
\newacronym{rrm}{RRM}{Radio Resource Management}
\newacronym{rru}{RRU}{Remote Radio Unit}
\newacronym{rs}{RS}{Remote Server}
\newacronym{rsrq}{RSRQ}{Reference Signal Received Quality}
\newacronym{rss}{RSS}{Received Signal Strength}
\newacronym{rssi}{RSSI}{Received Signal Strength Indicator}
\newacronym{rtt}{RTT}{Round Trip Time}
\newacronym{ru}{RU}{Radio Unit}
\newacronym{rus}{RSU}{Road Side Unit}
\newacronym{rw}{RW}{Receive Window}
\newacronym{rx}{RX}{Receiver}
\newacronym{s1ap}{S1AP}{S1 Application Protocol}
\newacronym{sa}{SA}{standalone}
\newacronym{sack}{SACK}{Selective Acknowledgment}
\newacronym{sap}{SAP}{Service Access Point}
\newacronym{sc2}{SC2}{Spectrum Collaboration Challenge}
\newacronym{scef}{SCEF}{Service Capability Exposure Function}
\newacronym{sch}{SCH}{Secondary Cell Handover}
\newacronym{scoot}{SCOOT}{Split Cycle Offset Optimization Technique}
\newacronym{sctp}{SCTP}{Stream Control Transmission Protocol}
\newacronym{sdap}{SDAP}{Service Data Adaptation Protocol}
\newacronym{sdk}{SDK}{Software Development Kit}
\newacronym{sdm}{SDM}{Space Division Multiplexing}
\newacronym{sdma}{SDMA}{Spatial Division Multiple Access}
\newacronym{sdn}{SDN}{Software-defined Networking}
\newacronym{sdr}{SDR}{Software-defined Radio}
\newacronym{seba}{SEBA}{SDN-Enabled Broadband Access}
\newacronym{sgsn}{SGSN}{Serving GPRS Support Node}
\newacronym{sgw}{SGW}{Service Gateway}
\newacronym{si}{SI}{Study Item}
\newacronym{sib}{SIB}{Secondary Information Block}
\newacronym{sinr}{SINR}{Signal to Interference plus Noise Ratio}
\newacronym{sip}{SIP}{Session Initiation Protocol}
\newacronym{siso}{SISO}{Single Input, Single Output}
\newacronym{sla}{SLA}{Service Level Agreement}
\newacronym{sm}{SM}{Service Model}
\newacronym{smo}{SMO}{Service Management and Orchestration}
\newacronym{smsgmsc}{SMS-GMSC}{\gls{sms}-Gateway}
\newacronym{snr}{SNR}{Signal-to-Noise-Ratio}
\newacronym{son}{SON}{Self-Organizing Network}
\newacronym{ngson}{NG SON}{Next Generation Self-Organizing Network}
\newacronym{sptcp}{SPTCP}{Single Path TCP}
\newacronym{srb}{SRB}{Service Radio Bearer}
\newacronym{srn}{SRN}{Standard Radio Node}
\newacronym{srs}{SRS}{Sounding Reference Signal}
\newacronym{ss}{SS}{Synchronization Signal}
\newacronym{sss}{SSS}{Secondary Synchronization Signal}
\newacronym{st}{ST}{Spanning Tree}
\newacronym{svc}{SVC}{Scalable Video Coding}
\newacronym{tb}{TB}{Transport Block}
\newacronym{tcp}{TCP}{Transmission Control Protocol}
\newacronym{tdd}{TDD}{Time Division Duplexing}
\newacronym{tdm}{TDM}{Time Division Multiplexing}
\newacronym{tdma}{TDMA}{Time Division Multiple Access}
\newacronym{tfl}{TfL}{Transport for London}
\newacronym{tfrc}{TFRC}{TCP-Friendly Rate Control}
\newacronym{tft}{TFT}{Traffic Flow Template}
\newacronym{tgen}{TGEN}{Traffic Generator}
\newacronym{tip}{TIP}{Telecom Infra Project}
\newacronym{tm}{TM}{Transparent Mode}
\newacronym{to}{TO}{Telco Operator}
\newacronym{tr}{TR}{Technical Report}
\newacronym{trp}{TRP}{Transmitter Receiver Pair}
\newacronym{ts}{TS}{Technical Specification}
\newacronym{tti}{TTI}{Transmission Time Interval}
\newacronym{ttt}{TTT}{Time-to-Trigger}
\newacronym{tx}{TX}{Transmitter}
\newacronym{uas}{UAS}{Unmanned Aerial System}
\newacronym{uav}{UAV}{Unmanned Aerial Vehicle}
\newacronym{udm}{UDM}{Unified Data Management}
\newacronym{udp}{UDP}{User Datagram Protocol}
\newacronym{udr}{UDR}{Unified Data Repository}
\newacronym{ue}{UE}{User Equipment}
\newacronym{uhd}{UHD}{\gls{usrp} Hardware Driver}
\newacronym{ul}{UL}{Uplink}
\newacronym{um}{UM}{Unacknowledged Mode}
\newacronym{uml}{UML}{Unified Modeling Language}
\newacronym{upa}{UPA}{Uniform Planar Array}
\newacronym{upf}{UPF}{User Plane Function}
\newacronym{urllc}{URLLC}{Ultra Reliable and Low Latency Communications}
\newacronym{usa}{U.S.}{United States}
\newacronym{usim}{USIM}{Universal Subscriber Identity Module}
\newacronym{usrp}{USRP}{Universal Software Radio Peripheral}
\newacronym{utc}{UTC}{Urban Traffic Control}
\newacronym{vim}{VIM}{Virtualization Infrastructure Manager}
\newacronym{vm}{VM}{Virtual Machine}
\newacronym{vnf}{VNF}{Virtual Network Function}
\newacronym{volte}{VoLTE}{Voice over \gls{lte}}
\newacronym{voltha}{VOLTHA}{Virtual OLT HArdware Abstraction}
\newacronym{vr}{VR}{Virtual Reality}
\newacronym{vran}{vRAN}{Virtualized \gls{ran}}
\newacronym{vss}{VSS}{Video Streaming Server}
\newacronym{v2x}{V2X}{Vehicle-to-Everything}
\newacronym{v2i}{V2I}{Vehicle-to-Infrastructure}
\newacronym{v2v}{V2V}{Vehicle-to-Vehicle}
\newacronym{cv2v}{C-V2V}{Cellular-\gls{v2v}}
\newacronym{cv2x}{C-V2X}{Cellular-V2X}
\newacronym{v2n}{V2N}{vehicle-to-network}
\newacronym{wbf}{WBF}{Wired Bias Function}
\newacronym{wf}{WF}{Waterfilling}
\newacronym{wg}{WG}{Working Group}
\newacronym{wlan}{WLAN}{Wireless Local Area Network}
\newacronym{osm}{OSM}{Open Source \gls{nfv} Management and Orchestration}
\newacronym{pnf}{PNF}{Physical Network Function}
\newacronym{drl}{DRL}{Deep Reinforcement Learning}
\newacronym{mtc}{MTC}{Machine-type Communications}
\newacronym{osc}{OSC}{O-RAN Software Community}
\newacronym{mns}{MnS}{Management Services}
\newacronym{ves}{VES}{\gls{vnf} Event Stream}
\newacronym{ei}{EI}{Enrichment Information}
\newacronym{fh}{FH}{Fronthaul}
\newacronym{fft}{FFT}{Fast Fourier Transform}
\newacronym{laa}{LAA}{Licensed-Assisted Access}
\newacronym{plfs}{PLFS}{Physical Layer Frequency Signals}
\newacronym{ptp}{PTP}{Precision Time Protocol}
\newacronym{lidar}{LiDAR}{Light Detection And Ranging}
\newacronym{dem}{DEM}{Digital Elevation Model}
\newacronym{dtm}{DEM}{Digital Terrain Model}
\newacronym{dsm}{DEM}{Digital Surface Models}
\newacronym{ota}{OTA}{Over-The-Air}
\newacronym{ns}{NS}{Network Slicing}
\newacronym{ne}{NE}{Nash Equilibrium}
\newacronym{hf}{HF}{High Frequency}
\newacronym{noma}{NOMA}{Non-Orthogonal Multiple Access}
\newacronym{sre}{SRE}{Smart Radio Environment}
\newacronym{ris}{RIS}{Reconfigurable Intelligent Surface}
\newacronym{inp}{InP}{Infrastructure Provider}
\newacronym{smf}{SMF}{Slicing Magangement Framework}
\newacronym{nsn}{NSN}{Network Slicing Negotiation}
\newacronym{sms}{SMS}{Slicing MAC Scheduler}
\newacronym{brd}{BRD}{Best Response Dynamics}
\newacronym{dssbr}{DSSBR}{Double Step Smoothed Best Response}
\newacronym{poa}{PoA}{Price of Anarchy}
\newacronym{pos}{PoS}{Price of Stability}
\newacronym{milp}{MILP}{Mixed Integer-Linear Program}
\newacronym{pod}{PoD}{Price of DSSBR}
\newacronym{roc}{ROC}{Radio Overload Control}
\newacronym{ciot}{cIoT}{critical Internet of Things}
\newacronym{embbpr}{eMBB Pr.}{enhanced Mobile BroadBand Premium}
\newacronym{sps}{SPS}{Semi-persistent Scheduling}
\newacronym{cg}{CG}{Configured Grant}
\newacronym{embbbs}{eMBB Bs.}{enhanced Mobile BroadBand Basic}
\newacronym{en}{EN}{Edge Node}
\newacronym{ec}{EC}{Edge Computing}
\newacronym{sp}{SP}{Service Provider}
\newacronym{me}{ME}{Market Equilibrium}
\newacronym{so}{SO}{Social Optimum}
\newacronym{wso}{WSO}{Weighted Social Optimum}
\newacronym{ps}{PS}{Proportional Sharing}
\newacronym{eg}{EG}{Eisenberg-Gale program}
\newacronym{pe}{PE}{Pareto Efficiency}
\newacronym{nsw}{NSW}{Nash Social Welfare}
\newacronym{ef}{EF}{Envy-Freeness}
\newacronym{sub6}{sub-6GHz}{Below 6GHz}
\newacronym{ncr}{NCR}{Network-Controlled Repeater}
\newacronym{nlos}{NLoS}{Non-Line of Sight}
\newacronym{src}{SRC}{Smart Radio Connection}
\newacronym{srd}{SRD}{Smart Radio Device}
\newacronym{cs}{CS}{Candidate Site}
\newacronym{tp}{TP}{Test Point}
\newacronym{fov}{FoV}{Field of View}
\newacronym{nrric}{near-RT RIC}{Near Real-time {RAN} Intelligent Controller}
\newacronym{e2ap}{E2AP}{E2 Application Protocol}
\newacronym{e2sm}{E2SM}{E2 Service Model}
\newacronym{nrtric}{non-RT RIC}{Non-Real-Time {RIC}}
\newacronym{itti}{ITTI}{Inter-task Interface}
\newacronym{bap}{BAP}{Backhaul Adaptation Protocol}
\newacronym{iabest}{IABEST}{Integrated Access and Backhaul Experimental large-Scale Tetbed}
\newacronym{teid}{TEID}{Tunnel Endpoint Identifier}
\newacronym{dlsch}{DL-SCH}{Downlink Shared Channel }
\newacronym{ulsch}{UL-SCH}{Uplink Shared Channel }
\newacronym{rsu}{RSU}{Road Side Unit}
\newacronym{its}{ITS}{Intelligent Transportation Systems}
\newacronym{vanet}{VANET}{Vehicular Ad-hoc Network}
\newacronym{dt}{DT}{Digital Twin}
\newacronym{ecc}{ECC}{Edge Computing Cluster}
\newacronym{obu}{OBU}{On Board Unit}
\newacronym{prdr}{PRDR}{Packet Reception Disagreement Ratio}
\newacronym{dr}{DR}{Disagreement Ratio}
\newacronym{ndt}{NDT}{Network Digital Twin}
\newacronym{cam}{CAM}{Cooperative Awareness Message}
\newacronym{cpm}{CPM}{Collective Perception Message}
\newacronym{pdf}{PDF}{Probability Density Function}
\newacronym[
  firstplural=Radio Access Technologies (RATs),
  longplural=Radio Access Technologies
]{rat}{RAT}{Radio Access Technology}
\newacronym{nlosv}{NLoSv}{Non-Line of Sight due to Vehicles}
\newacronym{gnss}{GNSS}{Global Navigation Satellite System}
\newacronym{rtk}{RTK}{Real-Time Kinematic}
\newacronym{sbr}{SBR}{Shooting and Bouncing Rays}

\modulolinenumbers[5]
\usepackage[utf8]{inputenc}
\usepackage{tikz}
\usepackage{pgfplots}
\usepackage{tikz-3dplot}
\usetikzlibrary{calc,patterns,angles,quotes}
\tdplotsetmaincoords{60}{115}
\pgfplotsset{compat=newest}
\usepackage{tabularx}
\usepackage[export]{adjustbox}

\newcommand{\firstrev}[1]{\textcolor{black}{#1}}
\newcommand{\secondrev}[1]{{\color{black}{#1}}}

\begin{document}

\title{VaN3Twin: the Multi-Technology V2X Digital Twin \\ with Ray Tracing in the Loop}

\author{
Roberto Pegurri\IEEEauthorrefmark{1}, Diego Gasco\IEEEauthorrefmark{2}, Francesco Linsalata\IEEEauthorrefmark{1}, \\ Marco Rapelli\IEEEauthorrefmark{2}, Eugenio Moro\IEEEauthorrefmark{1}, Francesco Raviglione\IEEEauthorrefmark{3}, Claudio Casetti\IEEEauthorrefmark{2}\\[0.2cm]

\IEEEauthorrefmark{1}Department of Electronics, Information and Bioengineering, Politecnico di Milano, Italy\\ 
\IEEEauthorrefmark{2}Department of Control and Computer Engineering, Politecnico di Torino, Italy\\
\IEEEauthorrefmark{3}Department of Electronics and Telecommunications, Politecnico di Torino, Italy\\[0.2cm]

\{roberto.pegurri, francesco.linsalata, eugenio.moro\}@polimi.it

\{diego.gasco, marco.rapelli, francesco.raviglione, claudio.casetti\}@polito.it}

\maketitle

\begin{abstract}
This paper presents VaN3Twin---the first open source, full-stack Network Digital Twin (NDT) framework for simulating the coexistence of multiple Vehicle-to-Everything (V2X) communication technologies with accurate physical-layer modeling via ray tracing. VaN3Twin extends the ms-van3t simulator by integrating Sionna Ray Tracer (RT) in the loop, enabling high-fidelity representation of wireless propagation, including diverse Line of Sight (LoS) conditions with a focus on LoS blockage due to other vehicles' meshes, Doppler effect, and site-dependent effects---e.g., scattering and diffraction. Unlike conventional simulation tools, the proposed framework supports realistic coexistence analysis across Dedicated Short Range Communication (DSRC) and Cellular-V2X (C-V2X) technologies operating over a shared spectrum. A dedicated interference tracking module captures cross-technology interference at the time-frequency resource block level and enhances Signal to Interference plus Noise Ratio (SINR) estimation by eliminating artifacts such as the bimodal behavior induced by separate LoS/NLoS propagation models. Compared to field measurements, VaN3Twin reduces application-layer disagreement by 50\% in rural and over 70\% in urban environments with respect to current state of the art simulation tools, demonstrating its value for scalable and accurate digital twin–based V2X coexistence simulation.
\end{abstract}


\section{Introduction}
\label{introduction}



In 2016, Toyota was the first car manufacturer to release commercial vehicles with \gls{v2x} communication capabilities, using \gls{dsrc} based on IEEE 802.11p for the Japanese market~\cite{Abuelsamid2018Toyota}. Shortly thereafter, similar deployments appeared in the U.S. and Europe~\cite{masini2018survey,5g2017assessment}, with General Motors and Volkswagen Group respectively. Today, the \gls{v2x} market continues to grow, with an increasing number of vehicles capable of communicating with thousands of \glspl{rsu} deployed along major highways~\cite{croads2024webinar}.
From a technological perspective, however, the landscape has grown increasingly complex. Since the introduction of LTE-V2X in 3GPP Rel. 14 (2017), Toyota has steered the Asian \gls{v2x} market toward \gls{cv2x} technology. In contrast, Europe continues to rely primarily on IEEE 802.11p for \gls{v2x} communication. Nevertheless, major auto-makers such as Audi, Mercedes, and Ford have frequently announced plans for future deployments leveraging \gls{cv2x} technologies (i.e., LTE-V2X and the emerging NR-V2X standard)~\cite{lucarelli2023auto, euro2022mercedes, ford2019ford}.
As a result, vehicles from different manufacturers---each potentially using distinct V2X technologies---will soon coexist on the road, creating significant challenges for interoperability, standardization, and coordinated deployment. In anticipation of this scenario, and as evidence of its growing relevance, many \gls{obu} and \gls{rsu} manufacturers have started equipping their products with support for multiple V2X technologies. However, the coexistence of these technologies remains a significant challenge. In the United States, the FCC reallocated 45 MHz of the 5.9 GHz safety band to unlicensed services and reserved the remaining 30 MHz exclusively for C-V2X, with follow-up rules adopted in December 2024 instructing legacy DSRC installations to migrate. Instead, in Europe both C-V2X and IEEE 802.11p operate within the same 5.8–5.9~GHz frequency band, therefore the issue of co-channel interference has emerged as a critical area of research in the V2X field. Indeed, numerous researchers and vehicular consortia are actively investigating these challenges~\cite{bazzi2020co, roux2020performance, c2ccc-coexistence}, which have emerged as some of the most pressing and widely discussed topics in the \gls{v2x} domain.
Predicting the performance of C-V2X and DSRC technologies in real-world coexistence scenarios is highly complex, especially in large-scale urban environments with fleets of vehicles equipped with either or both communication technologies. Although recent advancements in emulation frameworks for vehicular field testing~\cite{rapelli2024oscar} have significantly enhanced evaluation capabilities, conducting a comprehensive coexistence and co-channel interference study in real-world scenarios would require a large fleet of V2X-equipped vehicles capable of supporting one or more V2X \glspl{rat}. However, current OBU offerings from vendors typically support only a single \gls{rat} at a time---e.g., either DSRC or C-V2X, but not both simultaneously. This limitation makes large-scale field testing for coexistence scenarios logistically, economically, and technologically challenging. As a result, advanced simulation tools are essential to analyze the complex, large-scale scenarios expected in future vehicular networks. Moreover, accurate interference modeling is crucial to support research on agile spectrum utilization and effective spectrum sharing across heterogeneous radio technologies.
Beyond traditional simulation goals, high-fidelity modeling is also a cornerstone of the emerging paradigm of \gls{ndt}~\cite{roongpraiwan2025digital}. An \gls{ndt} aims to create an accurate replica in real time of a network, requiring simulation models to maintain a high level of precision in all layers of the communication stack~\cite{Pegu2505:Toward}. This requirement is particularly critical in the context of \gls{v2x} wireless networks, where operating conditions evolve rapidly due to high mobility, complex propagation environments---especially in urban scenarios---and finally, multi-RAT coexistence. In such settings, and depending on the final scope of the simulation, precision becomes indispensable to ensure that the digital replica faithfully mirrors its physical counterpart.
As is often the case, the more realistic the simulator, the more sophisticated and reliable the resulting models. However, to date, few vehicular simulation tools incorporate realistic channel models. For instance, simulators based on Network Simulator Version 3 (ns-3)~\cite{ns-3-book-chapter} often rely on stochastic channel models that omit critical features, such as deterministic shadowing and multipath effect characterization and comprehensive propagation loss models. Another critical limitation, extremely relevant in V2V scenarios, is their failure to account for the impact of surrounding vehicles on the received signal.
Furthermore, only a limited number of vehicular simulators have native support for full-stack implementation of multiple communication technologies, including IEEE 802.11p, LTE-V2X, and NR-V2X, and integration with the higher layers standardized by ETSI. These limitations have hindered the development of robust co-channel interference studies using vehicular simulations. Addressing these gaps is crucial for advancing our understanding of coexistence scenarios and improving the reliability of simulation-based research.

In this article, we present VaN3Twin\footnote{\url{https://github.com/DriveX-devs/VaN3Twin}}, a framework based on the integration of Sionna Ray Tracer (RT)~\cite{sionna}, an open source\footnote{https://github.com/nvlabs/sionna} ray tracing engine for high-fidelity wireless channel modeling, into the ms-van3t V2X simulator~\cite{ms-van3t-journal-2024}, resulting in a reliable and \gls{ndt}-ready framework for full-stack and multi-RAT V2X simulations. VaN3Twin also has built-in support for modeling the coexistence of multiple communication technologies and intrinsic consideration for both static (buildings, trees, and roadside infrastructure) and dynamic objects (vehicles, pedestrians, etc.) in the simulated propagation environment. The ms-van3t simulator represents the state of the art among open source V2X simulators built on the ns-3 platform. It supports the design, simulation, and hardware-in-the-loop emulation of ETSI-compliant V2X applications. 
However, as ns-3 native simulator, ms-van3t traditionally relies on probabilistic models to simulate wireless channel effects. Although effective for large-scale simulations, these models often oversimplify physical-layer phenomena, limiting accuracy in dynamic and interference-prone environments. By using ray tracing, VaN3Twin overcomes this limitation, bridging the gap between high-level vehicular simulation and realistic modeling. 

Our main contributions are summarized as follows.
\begin{itemize} 
    \item Realization of a first-of-its-kind open source, full-stack, and multi-RAT V2X \gls{ndt} named VaN3Twin, overcoming the limitations of traditional stochastic and frequency-dependent channel models with ray tracing. \firstrev{While \cite{Pegu2505:Toward} focused exclusively on the integration of Sionna RT into the ns-3 loop, the present work introduces a complete V2X \gls{ndt}, focusing on accurate simulation of scenarios of multiple V2X technologies sharing the same spectrum, thereby explicitly considering their coexistence;}
    
    \item Development of an adaptable and modular framework, allowing the ray tracing component to run on a dedicated GPU-equipped server, while the rest of VaN3Twin may be executed on a separate machine. This division not only optimizes computational resources but also increases scalability for high-fidelity, city-scale V2X simulations;
    
    \item Validation of VaN3Twin as a \gls{ndt}, performed against field measurements in real-world rural and urban testbeds. The results demonstrate significant improvements in fidelity at both the propagation and packet reception levels, 
    with VaN3Twin reducing the disagreement with field data by 50\% in rural and over 70\% in urban environments when compared to state of the art simulation tools;
    
    \item Development of a unified coexistence modeling framework capable of simulating heterogeneous deployments with one or more \glspl{rat} simultaneously within a shared spectrum portion. This includes a dedicated VaN3Twin coexistence module that incorporates the effects of co-channel coexistence into \gls{sinr} estimation;
    
    \item Demonstration of improved \gls{sinr} estimation accuracy in coexistence scenarios, showing that ray-based simulations produce smoother SINR distributions and eliminate the artificial bimodality introduced by the use of separate propagation models for \gls{los} and \gls{nlos} conditions.
\end{itemize}

The remainder of the article is organized as follows.
Section II details VaN3Twin, highlighting its \firstrev{flexible} architecture, the integration of a ray tracer, data exchange mechanisms, and mobility synchronization strategies.
Section III validates the accuracy of the proposed framework's ray-based propagation model by comparing simulation results with real-world vehicular measurements in rural and urban environments, highlighting improved alignment in both \gls{rssi} values and packet reception performance.
Section IV introduces the coexistence module for multi-\gls{rat} interference management in ms-van3t and VaN3Twin. It presents a simulation comparing SINR estimation under 3GPP-based and ray-based models in co-channel scenarios, with particular emphasis on \gls{nlos} conditions caused by vehicular obstructions, and evaluates the resulting impact on cumulative and punctual packet-level performance. \firstrev{The section also provides an analysis on VaN3Twin scalability.}
Section V reviews related work on V2X coexistence simulation and ray tracing integration.
Finally, Section VI concludes the paper. 

\section{VaN3Twin Framework}\label{sec:integration}

\begin{figure}[!t]
    \includegraphics[width=1.0\linewidth]{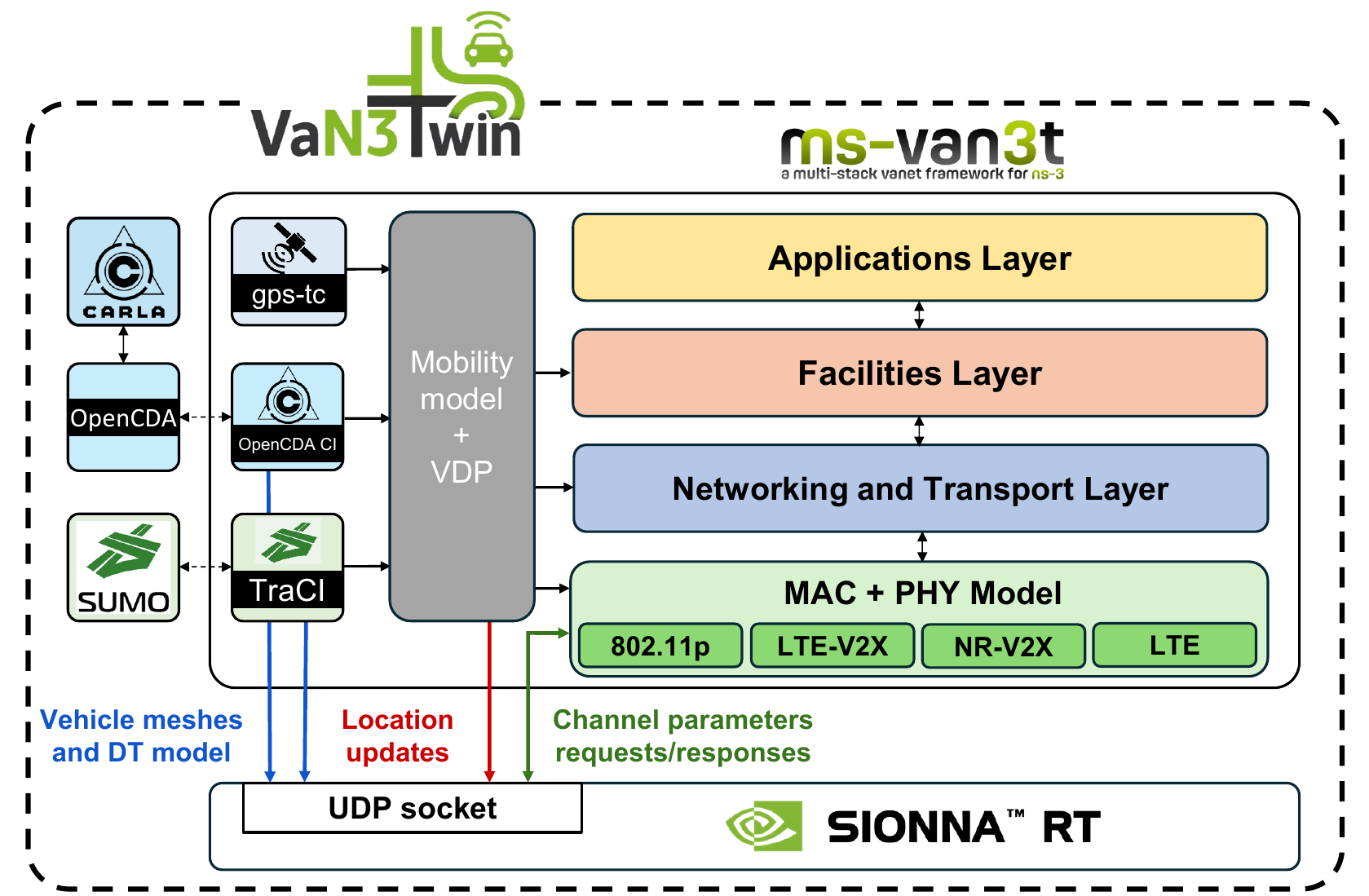}
\caption{An overview of the proposed VaN3Twin framework: ms-van3t high-level architecture with Sionna RT integration.} 
\label{fig:ms-van3t-arch-Sionna}
\end{figure}

This section presents the proposed architecture, which integrates Sionna RT into the existing \mbox{ms-van3t} workflow, as illustrated in Figure~\ref{fig:ms-van3t-arch-Sionna}. Specifically, the figure displays the complete ms-van3t stack, spanning from the Application Layer to the Physical Layer---the latter expanded to show all the available \glspl{rat} in ms-van3t. Figure~\ref{fig:ms-van3t-arch-Sionna} also outlines the mobility interfaces of ms-van3t, which can interact with either SUMO~\cite{SUMO2018} or CARLA~\cite{dosovitskiy2017carla} mobility simulators or with pre-recorded vehicular traces. When using SUMO or CARLA, the mobility state of each vehicle is retrieved through standardized V2X message exchange interfaces, namely the Traffic Control Interface (TraCI)~\cite{traci} for SUMO and OpenCDA for CARLA. Mobility data are eventually received by the ms-van3t Mobility model and subsequently distributed to the appropriate layers via the Vehicle Data Provider (VDP) module. Finally, at the bottom of Figure~\ref{fig:ms-van3t-arch-Sionna}, the integration with Sionna RT is shown. This integration is achieved through data exchange via a UDP socket, enabling a fast and modular architecture. As a result, the ray tracer can operate either on the same local machine as ms-van3t or on a dedicated remote server with enhanced computational capabilities, \firstrev{as modularity is a key feature of VaN3Twin.}
\firstrev{Furthermore, by relying on ms-van3t and its underlying ns-3 engine, every module of VaN3Twin can be either replaced, extended, or re-implemented from scratch, including protocol stacks, channel models, mobility patterns, and performance metrics. Moreover, the employed ray tracing engine (Sionna RT) is open source and fully customizable. Through a dedicated adapter interface, it can be entirely substituted without affecting the integration logic. This design choice maximizes flexibility for secondary development---following standard \mbox{ns-3} extension practices, such as models, helpers, and attributes---and facilitates the adoption of future improvements in ray tracing technology. Documentation\footnote{https://www.nsnam.org/documentation} and APIs are provided to guide developers, thereby supporting secondary development and experimentation.}
As for the information exchange between ms-van3t and the ray tracer engine, the communication is structured into three distinct data flows, each identified by matching color codes in Figures~\ref{fig:ms-van3t-arch-Sionna} and Figure~\ref{fig:flowchart}, where Figure~\ref{fig:flowchart} offers a more detailed representation of the interaction workflow. Specifically, the three data flows comprise: (i) the initialization of the \gls{dt} \firstrev{including buildings and roadside infrastructure} and the mapping of vehicle meshes, represented in the figures by blue arrows; (ii) the transmission of dynamic location updates, indicated in red; (iii) the request and response exchange of channel parameters, depicted in green. Each of these flows is analyzed in greater detail in the following subsections.

\subsection{DT model initialization and vehicle mesh mapping}
\label{subsec:blueflow}
\begin{figure}[!b]
    \centering
    \includegraphics[width=1\linewidth]{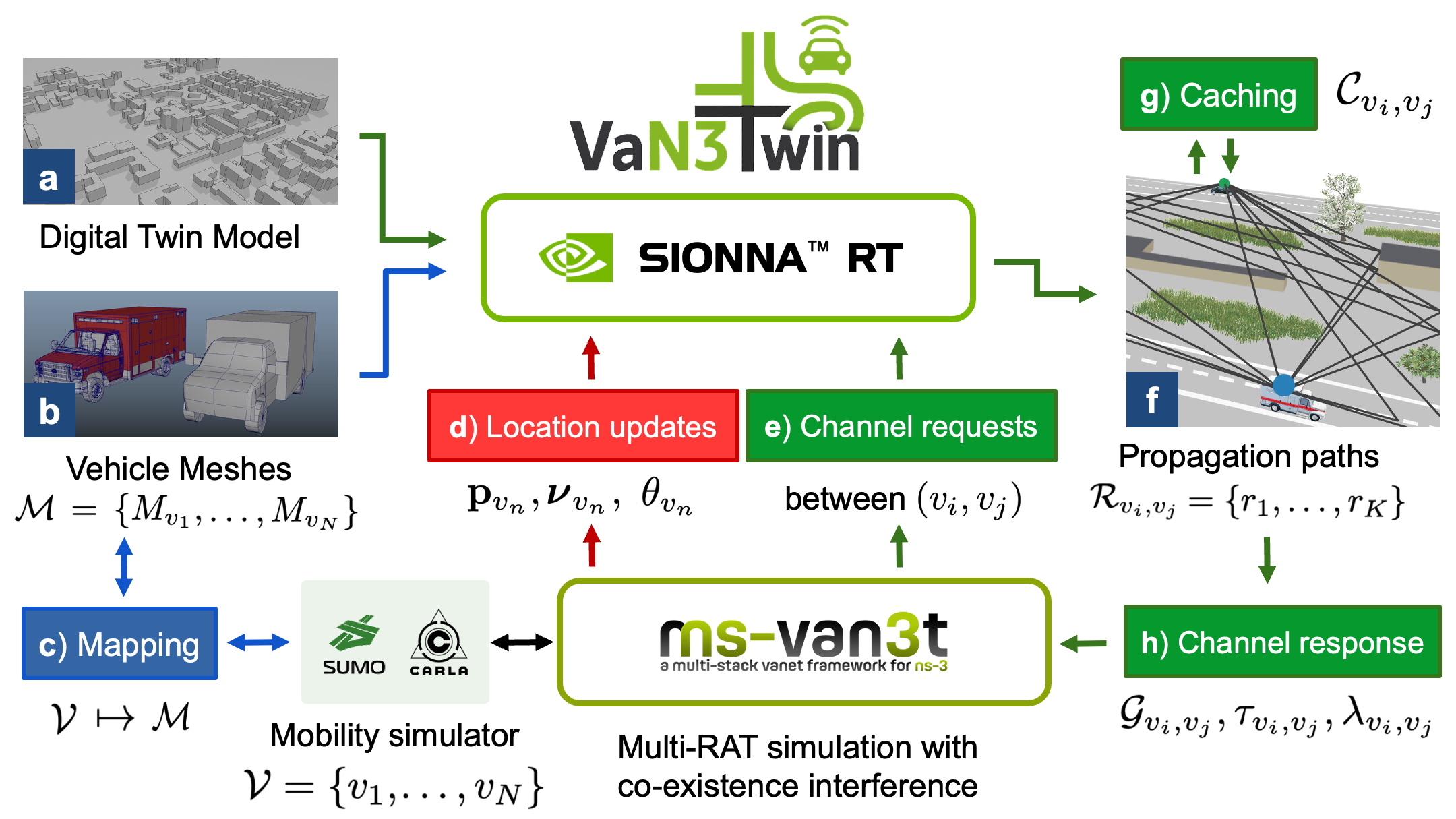}
    \caption{Flowchart of the entire VaN3Twin framework.}
    \label{fig:flowchart}
\end{figure}

The ray tracer operates within a Three Dimensional (3D) scenario that includes both static and dynamic objects, each represented by a 3D mesh. The geometry of each mesh defines the physical shape of the object, while its material properties (e.g., metal, glass, or plastic) govern its electromagnetic characteristics, thereby influencing the interaction of electromagnetic waves with the object's surface. 
Static objects (such as buildings, road signs, and roadside units) remain fixed throughout the simulation and constitute our DT model. 
To support the simulation of vehicular traffic within the modeled environment, the scenario topology---including polygons for buildings---was extracted from OpenStreetMap (OSM)~\cite{osm} and imported into the SUMO traffic simulator. This integration enables the construction of a complete DT model of the infrastructures in our scenario (as illustrated in Figure~\ref{fig:flowchart}a).
Conversely, dynamic objects represent mobile entities such as vehicles. For this purpose, various vehicle meshes (e.g., passenger cars, sedans, trucks, buses, etc.) were selected from the supported mobility simulators. Each vehicle mesh was further segmented into distinct components---i.e., vehicle body, lights, windows, rims, etc.---in order to assign appropriate radio material properties to each part. These vehicle mesh types were employed to model the dynamic elements of the simulation scenario. The process begins by defining the set of vehicles to be included in the overall simulation, defined as $\mathcal{V} = \{v_1, v_2, \ldots, v_N\}$. To distinguish vehicles participating in the simulation at time $t$, we define the subset of active vehicles $\mathcal{V}_\text{active}(t) \subseteq \mathcal{V}$. The set of 3D car meshes (shown in Figure~\ref{fig:flowchart}b) is denoted by $\mathcal{M} = \{M_{v_1}, M_{v_2}, \ldots, M_{v_N}\}$, where each mesh $M_{v_n} \in \mathcal{M}$ represents the geometry and material properties of vehicle $v_n$. Each active vehicle $v_n \in \mathcal{V}_\text{active}(t)$ is equipped with two antennas, $(a_{v_n, \mathrm{Tx}}, a_{v_n, \mathrm{Rx}})$, one for transmission and one for reception, mounted on the mesh $M_{v_n}$ at positions:
\begin{equation}
\mathbf{p}_{\text{ant}}(t') = \mathbf{p}_{v_n}(t') + \mathbf{p}_{\text{dis}},
\end{equation}
where $\mathbf{p}_{\text{dis}} = [x_{\text{dis}}, y_{\text{dis}}, z_{\text{dis}}]$ denotes the displacement of the antenna relative to reference position $\mathbf{p}_{v_n}(t')$ of vehicle $v_n$ at time $t' < t$, which is the time denoting the last update applied to the mesh $M_{v_n}$. Each vehicle $v_n \in \mathcal{V}$ is associated with a unique 3D mesh $M_{v_n} \in \mathcal{M}$ in the ray tracing scenario through a one-to-one mapping $\phi: v_n \mapsto M_{v_n}, \forall v_n \in \mathcal{V}$ (see Figure~\ref{fig:flowchart}c).
Both the DT initialization and the vehicle meshes mapping $\phi$ are initialized once at the beginning of the simulation and remain constant throughout, incurring no runtime computational cost. Whenever a vehicle $v_n$ becomes active---i.e., $v_n \in \mathcal{V}_\text{active}(t)$---the mesh $M_{v_n}$ is updated to reflect its physical state and its dynamic location updates.

\subsection{Dynamic location updates}
\label{subsec:redflow}

At simulation time $t$, each vehicle $v_n \in \mathcal{V}$ is characterized by a position vector $\mathbf{p}_{v_n}(t) = [x_{v_n}(t), y_{v_n}(t), z_{v_n}(t)]$, a velocity vector $\boldsymbol{\nu}_{v_n}(t) = [\nu_{x,v_n}(t), \nu_{y,v_n}(t), \nu_{z,v_n}(t)]$, and a heading angle $\theta_{v_n}(t)$. Moreover, at each simulation time step, information regarding the active vehicles---i.e., vehicle classes, vehicle IDs, positions, and orientations---is retrieved from the Mobility model of ms-van3t. To ensure a coherent and time-consistent representation of moving objects, dynamic location updates generated by the mobility simulator are transmitted from the Mobility model of ms-van3t to the ray tracer (Figure~\ref{fig:flowchart}d). Specifically, whenever an updated mobility state becomes available for a vehicle $v_n$ at time $t$, a location update message is sent to the ray tracer containing the tuple $\left\{\mathbf{p}_{v_n}(t),\, \boldsymbol{\nu}_{v_n}(t),\, \theta_{v_n}(t)\right\}$ related to vehicle $v_n$.
Upon reception, this information is integrated into the ray tracing scenario and used to update the radio propagation environment by repositioning the vehicle meshes according to the received positions and orientations, and by setting the corresponding vehicle configurations accordingly. Thus, at each update, the variables $\mathbf{p}_{v_n}(t)$, $\boldsymbol{\nu}_{v_n}(t)$, and $\theta_{v_n}(t)$ are updated to reflect the current state of each active vehicle $v_n \in \mathcal{V}_\text{active}(t)$ and each vehicle mesh $M_{v_n}$ is synchronized accordingly. The velocity $\boldsymbol{\nu}_{v_n}(t)$ is provided to allow for Doppler shift---and consequent channel coherence time---computation. This synchronization ensures that the propagation environment faithfully reflects the time-varying geometry and kinematics of the vehicular network, which is essential for the realistic modeling of wireless channel dynamics under mobility.
To limit the computational cost of ray tracing scene updates, synchronization is performed only if the vehicle has moved beyond a minimum threshold distance $\Delta d_{\text{min}}$, defined as:
\begin{equation}
\Delta d_{\text{min}} = \max \left( \Delta d_0,~ \nu_{v_n(t)} \cdot \tau_{\text{coh}} \right),
\end{equation}
where $\Delta d_0$ is a base distance threshold, $\tau_{\text{coh}}$ is the channel coherence time, and $\nu_{v_n(t)}$ the current velocity of the vehicle. Since a Doppler-based threshold may lead to excessive updates, vehicular simulations typically favor the use of larger $\Delta d_0$, \firstrev{with its numerical value being scenario-dependent and determined empirically.} \secondrev{Clearly, this simple approach may be extended into more advanced and efficient mechanism, also possibly linked to the specific geometry of the considered scenario.}
Dynamic location updates are applied only if the following condition holds:
\begin{equation}
\| \mathbf{p}_{v_n}(t) - \mathbf{p}_{v_n}(t') \| > \Delta d_{\text{min}},
\end{equation}
where $t' < t$ denotes the time of the last update applied to the mesh $M_{v_n}$. This condition is always true if a location update procedure for $M_{v_n}$ has never been completed before.

\subsection{Channel parameter requests and responses}
\label{subsec:greenflow}

This final data exchange, represented by the green arrows in  Figure~\ref{fig:ms-van3t-arch-Sionna} and Figure~\ref{fig:flowchart}, takes place between the ray tracing component and the Physical Layer of ms-van3t. Specifically, when the \gls{rat} model within ms-van3t detects an imminent transmission between two vehicles $(v_i, v_j) \in \mathcal{V}_{\text{active}}(t) \subseteq \mathcal{V}$ at simulation time $t$, a channel parameter request is issued to the ray tracer (see Figure~\ref{fig:flowchart}e), bypassing the native channel computation mechanisms of ns-3.
Upon receiving a channel parameter request, the ray tracer computes the set of propagation paths $\mathcal{R}_{v_i,v_j}$ (as illustrated in Figure~\ref{fig:flowchart}f) between the antennas mounted on the corresponding vehicle meshes $M_{v_i}$ and $M_{v_j}$. Although initially triggered by a specific vehicle pair, the ray tracing engine computes all pairwise propagation paths among the vehicles in $\mathcal{V}_\text{active}(t)$ to avoid redundant computations. This strategy, in conjunction with carefully selected trade-offs aimed at reducing ray tracing complexity while maintaining accuracy~\cite{9459462}, amortizes the cost of scene analysis \firstrev{by leveraging GPU parallelization and distributing initialization overhead across multiple channel queries}. The resulting set of admissible propagation paths, under the assumption of channel reciprocity, is:
\begin{equation}
\mathcal{R}_{v_i,v_j} = \mathcal{R}_{v_j,v_i} = \{ r_1, r_2, \ldots, r_K \}.
\end{equation}
Each path $r_k$ is characterized by a complex gain $\alpha_k$, a delay $\tau_k$, direction of departure $\boldsymbol{\psi}_k$, direction of arrival $\boldsymbol{\vartheta}_k$, and a Doppler shift $\varphi_k$. Only those paths that satisfy relevant physical and geometric constraints (e.g., interaction type, angular resolution, and bounce count) are included in $\mathcal{R}_{v_i,v_j}$. Depending on the characteristics of the environment, both \gls{los} and \gls{nlos} components may be present in the computed propagation paths.
In addition to the set of propagation paths $\mathcal{R}_{v_i,v_j}$, the channel impulse response $\mathbf{h}_{v_{i},v_{j}}(t)$ between a transmitting vehicle $v_{i}$ and a receiving vehicle $v_{j}$ is computed, derived using the impulse response model for multipath channels:
\begin{equation}\label{eq:channel_matrix_time}
    \mathbf{h}_{v_{i},v_{j}}(t) = \sum_{k=1}^{K}\alpha_k \, e^{j 2 \pi \varphi_k t}\,
    \mathbf{a}_{v_{i}}(\boldsymbol{\vartheta}_k)\mathbf{a}_{v_{j}}^{\mathrm{T}}(\boldsymbol{\psi}_k)\,
    g\left(t - \tau_k\right),
\end{equation}
where $\alpha_k$ denotes the complex path gain of the $k$-th ray, which depends on the underlying propagation mechanisms (i.e., reflection, diffraction, scattering), $\varphi_k$ represents the Doppler shift associated with the $k$-th path, resulting from the relative motion between $v_{i}$, and $v_{j}$, $\mathbf{a}_{v_{i}}(\boldsymbol{\vartheta}_k) \in \mathbb{C}^{N_T \times 1}$ and $\mathbf{a}_{v_{j}}^{\mathrm{T}}(\boldsymbol{\psi}_k) \in \mathbb{C}^{N_R \times 1}$ correspond to the transmit and transposed receive array response vectors, respectively, where $\boldsymbol{\psi}_k$ and $\boldsymbol{\vartheta}_k$ denote the directions of arrival and departure of the ray, $g(t - \tau_k)$ denotes the pulse shaping filter, with $\tau_k$ being the propagation delay associated with the $k$-th ray.
Among the various parameters involved in the channel impulse response $\mathbf{h}_{v_{i},v_{j}}(t)$ described above, a selected subset can be extracted and included in the channel parameter response sent to the Physical Layer of ms-van3t (see Figure~\ref{fig:flowchart}h). Specifically, each channel parameter response is composed of the tuple $\left\{\mathcal{G}_{v_i,v_j},\, \tau_{v_i,v_j},\, \lambda_{v_i,v_j}\right\}$, where $\mathcal{G}_{v_i,v_j}$ denotes the total path gain, defined as:
\begin{equation}
\mathcal{G}_{v_i,v_j} = \left| \sum_{k=1}^{K} \alpha_k \right|^2;
\end{equation}
$\tau_{v_i,v_j}$ denotes the propagation delay of the communication, defined as the minimum among all individual path delays, i.e.:
\begin{equation}
\tau_{v_i,v_j} = \min \left\{ \tau_1, \tau_2, \ldots, \tau_K \right\};
\end{equation}
\firstrev{where, $\tau_{v_i,v_j}$ is used solely to capture the end-to-end propagation delay, whereas the full set of multipath components is preserved in the channel characterization and fully incorporated in the subsequent receiver processing.}

$\lambda_{v_i,v_j}$ indicates the \gls{los} condition, and is defined as:
\begin{equation}
\lambda_{v_i,v_j} = 
\begin{cases}
1, & \text{if } \exists r_k \in \mathcal{R}_{v_i,v_j} \text{ that is LoS}, \\
0, & \text{otherwise}.
\end{cases}
\end{equation}

For computation efficiency reasons, the set of propagation paths $\mathcal{R}_{v_i,v_j}$ computed with ray tracing is stored in a cache $\mathcal{C}$ for potential reuse (shown in Figure~\ref{fig:flowchart}g). Each cache entry corresponding to a vehicle pair $(v_i, v_j)$ is denoted by $\mathcal{C}_{v_i,v_j}$. When a channel parameter request is issued at time $t$, two cases may arise:
\begin{itemize}
  \item \textit{Cache miss}: If $\mathcal{C}_{v_i,v_j} = \emptyset$. In this case, the ray tracing engine computes $\mathcal{R}_{v_i,v_j}$ for all $(v_i, v_j) \in \mathcal{V}_\text{active}(t)$, extracts the requested parameter, and stores the result:
 \mbox{$ \mathcal{C}_{v_i,v_j} \gets \mathcal{R}_{v_i,v_j}$};
 \item \textit{Cache hit}: If $\mathcal{C}_{v_i,v_j} \neq \emptyset$. In this case, the stored result is used without additional computation.
\end{itemize}


\firstrev{Whenever any vehicle $v_n$ undergoes a state change that results in an update of its associated mesh $M_{v_n}$, the entire cache for every vehicle pair $(v_i, v_j) \in \mathcal{V}_\text{active}(t)$ is invalidated: $\mathcal{C} \gets \emptyset$. This conservative strategy ensures cache consistency with the current simulation state, fully reflecting the dynamic impact of mobile scatterers (other vehicles) on propagation. While this mechanism may appear stringent, it is crucial in vehicular networks, where transient changes in the environment can substantially affect multipath structure, received power, and signal quality.} In the following section, we validate the ray tracing propagation model of VaN3Twin by comparing it against real-world vehicular field measurements.

\section{VaN3Twin Validation through Real-data} \label{sec:validation}


\begin{figure*} [!t]
    \centering
    \begin{subfigure}[t]{0.45\linewidth}
        \centering
        \includegraphics[width=\linewidth]{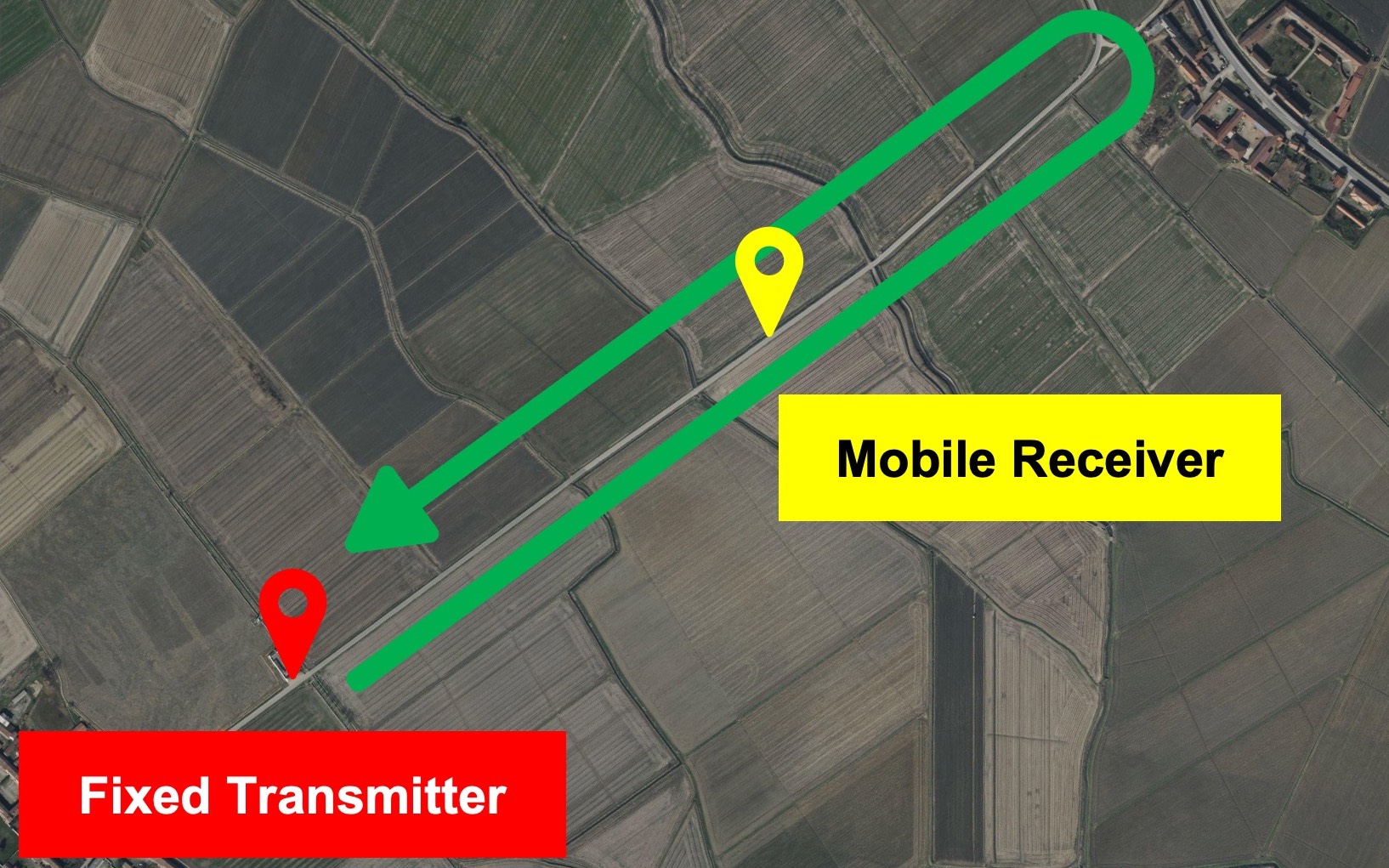}
        \caption{Satellite view of the rural scenario~\cite{maps}.}
        \label{fig:sali-sat}
        \hfill 
    \end{subfigure}
    \quad
    \begin{subfigure}[t]{0.45\linewidth}
        \centering
        \includegraphics[width=\linewidth]{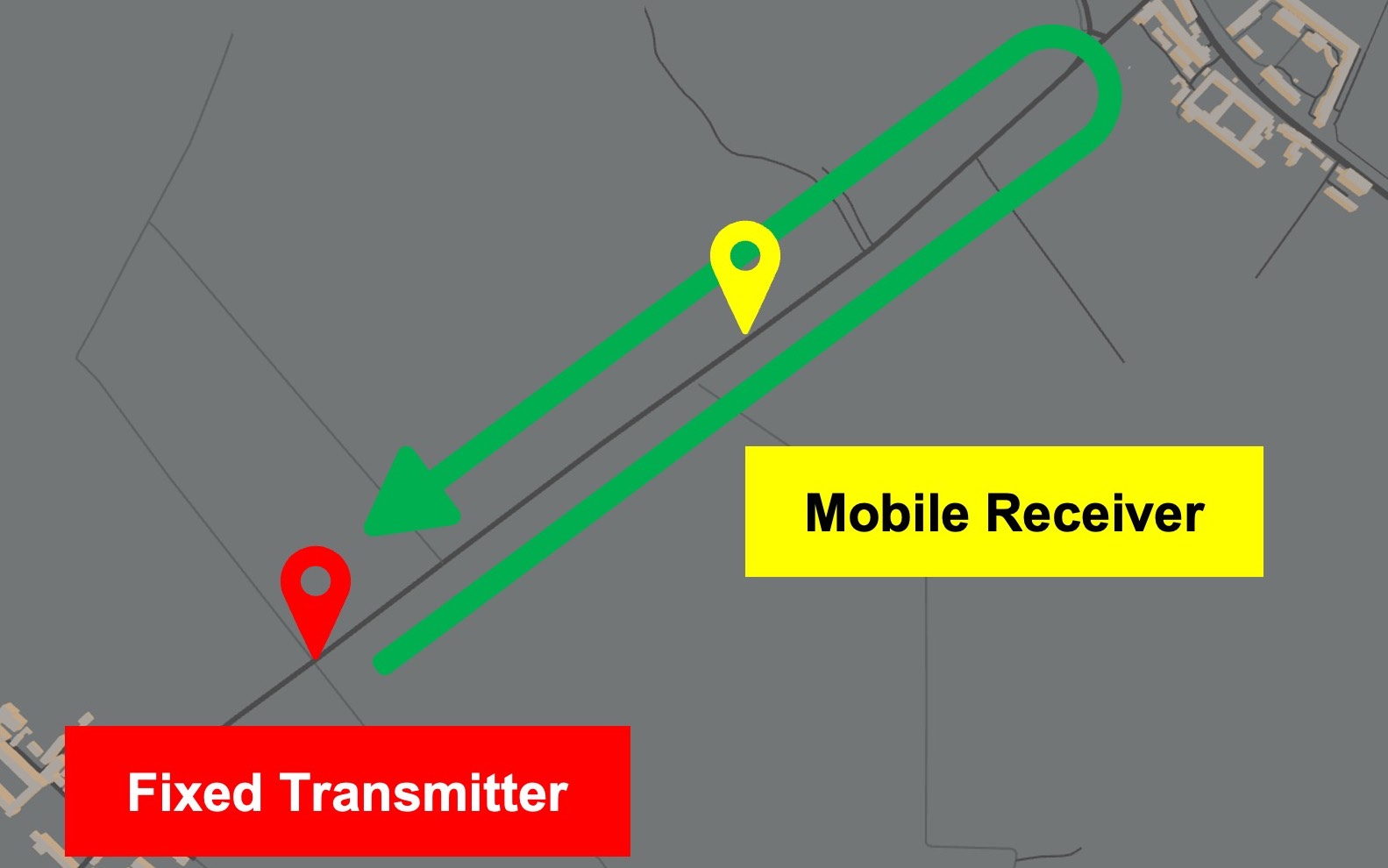}
        \caption{Replica for the VaN3Twin ray tracer of the rural scenario.}
        \label{fig:sali-sionna}
        \hfill
    \end{subfigure}
    \quad
    \begin{subfigure}[t]{0.45\linewidth}
        \centering
        \includegraphics[width=\linewidth]{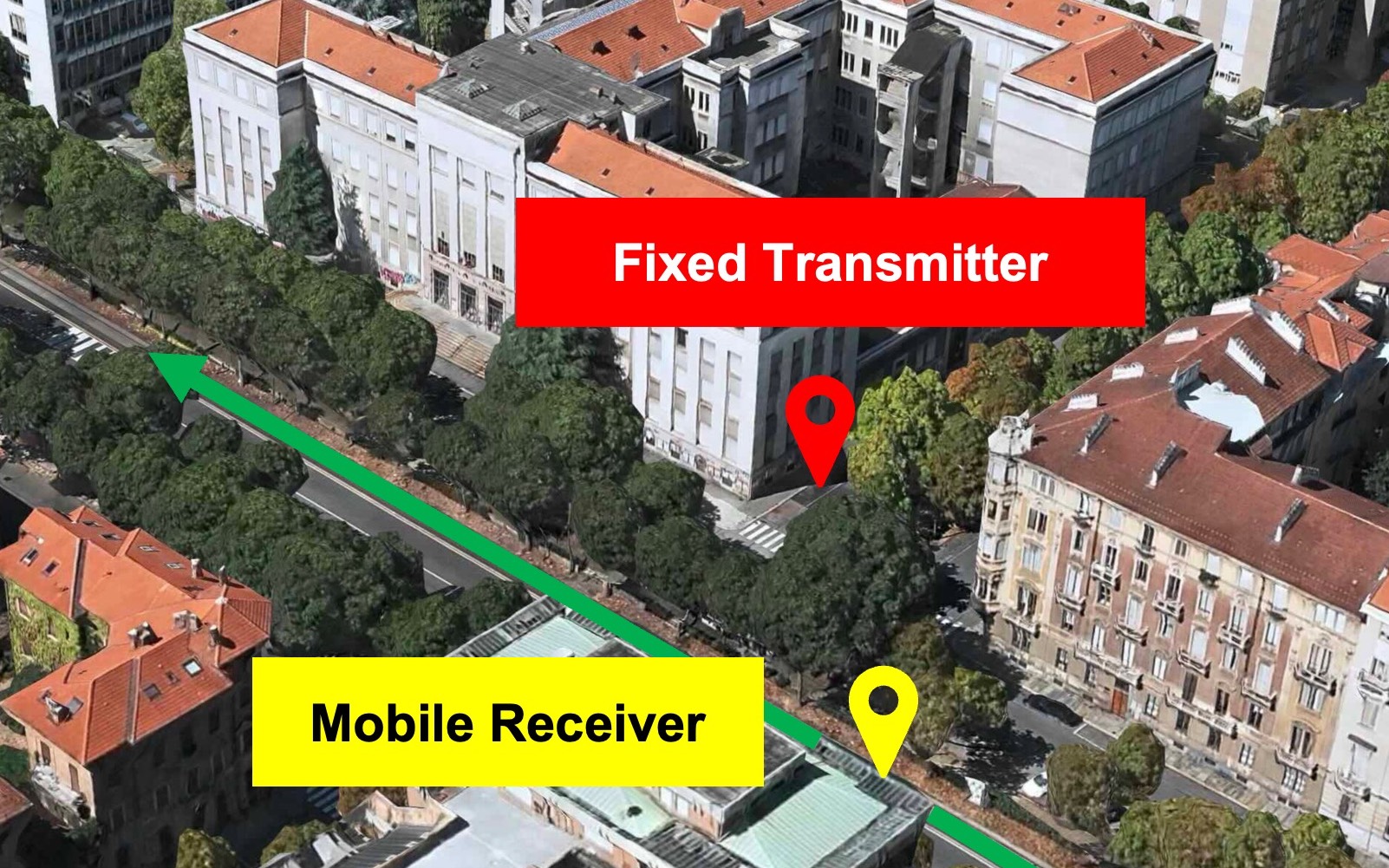}
        \caption{Satellite view of the urban scenario~\cite{maps}.}
        \label{fig:torino-sat}
        \hfill 
    \end{subfigure}
    \quad
    \begin{subfigure}[t]{0.45\linewidth}
        \centering
        \includegraphics[width=\linewidth]{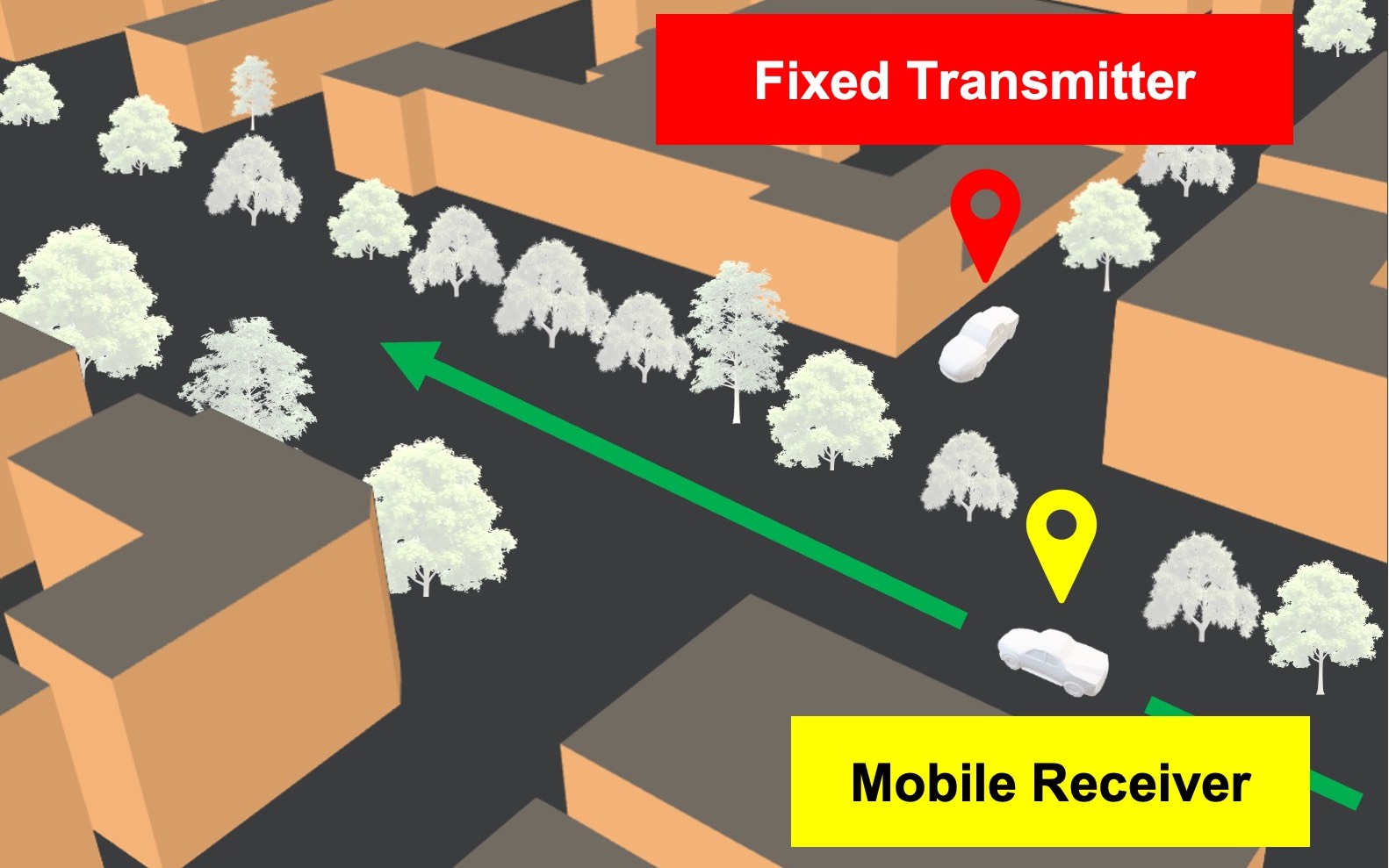}
        \caption{Replica for the VaN3Twin ray tracer of the urban scenario.}
        \label{fig:torino-sionna}
        \hfill 
    \end{subfigure}
    \caption{Areas of the two measurement campaigns in Sali Vercellese, Vercelli, Italy (a) and its DT (b) and Turin, Italy (c) and its DT (d), rural and urban scenarios, respectively. In both DTs, the static $T_x$ antenna is represented as the red marker, while the $R_x$ is yellow and moves along the green trajectory.}
    \label{fig:torino-scenari}
\end{figure*}

\begin{figure} [!b]
    \centering
    \begin{subfigure}[t]{0.45\linewidth}
        \centering
        \includegraphics[width=\linewidth]{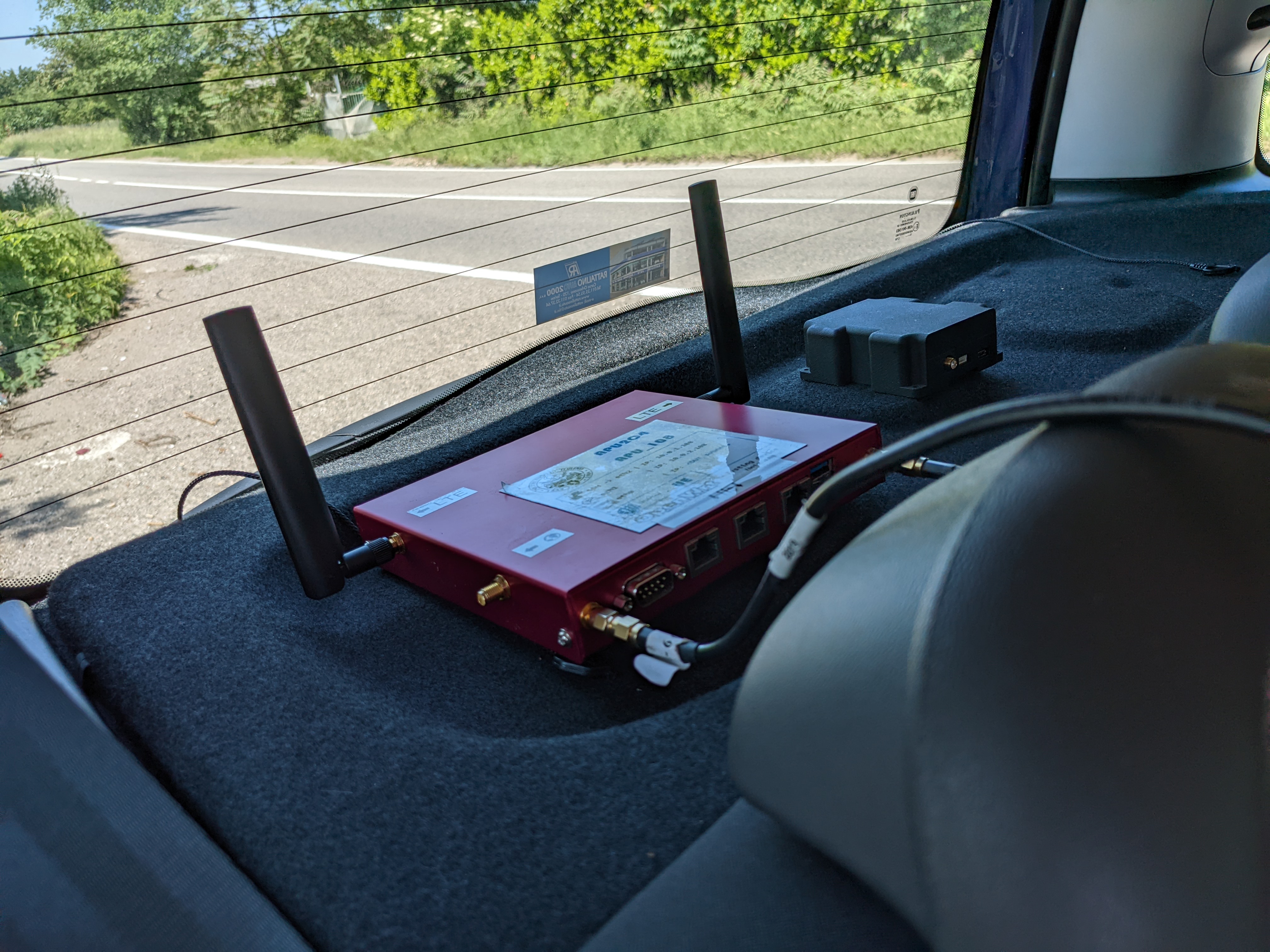}
        \caption{OBU and ArduSimple GNSS devices.}
        \label{fig:obu-test}
        \hfill 
    \end{subfigure}
    \quad
    \begin{subfigure}[t]{0.45\linewidth}
        \centering
        \includegraphics[width=\linewidth]{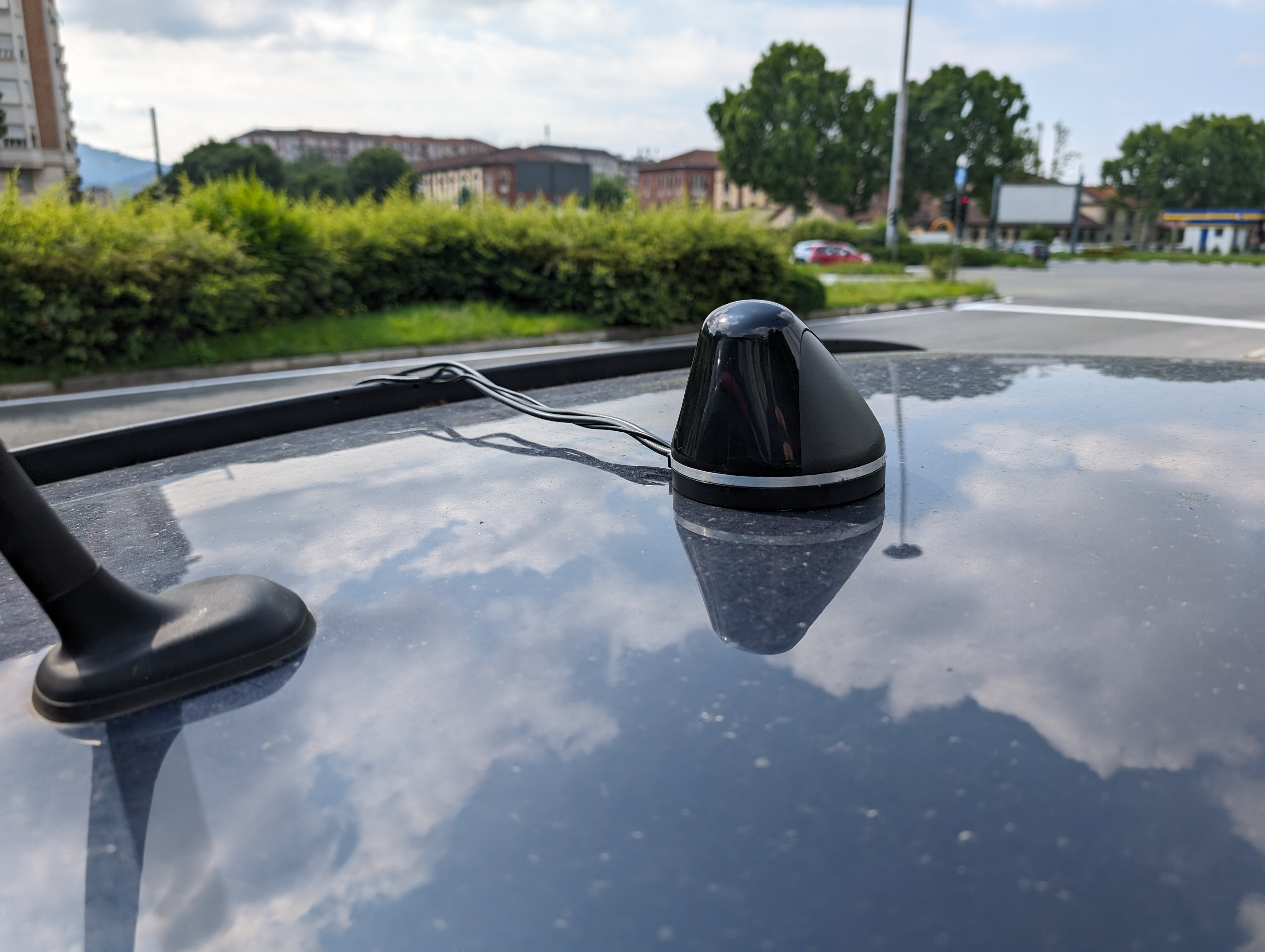}
        \caption{Magnetic MobileMark antenna.}
        \label{fig:antenna-test}
        \hfill
    \end{subfigure}
\caption{Vehicle setup for field tests.}
    \label{fig:fieldtest-setup}
\end{figure}

To assess how accurately VaN3Twin can replicate the wireless propagation in a V2X network, two measurement campaigns were conducted: one in a rural setting in Sali Vercellese, Vercelli, Italy (Figure~\ref{fig:sali-sat}) and the other in the city center of Turin, Italy (Figure~\ref{fig:torino-sat}).
The field tests were conducted using a dedicated setup, thoroughly detailed in~\cite{rapelli2024oscar} and illustrated in Figure~\ref{fig:fieldtest-setup}. In particular, Figure~\ref{fig:obu-test} shows an APU2E4 embedded board from PC Engines~\cite{apu2e4}---the red box in the figure---and an ArduSimple simpleRTK2B Fusion board~\cite{simpleRTK2BFusion}---the small black box. The OBU runs the latest version of OpenWrt-V2X~\cite{raviglione2019open} as its operating system, along with the most recent release of the Open Stack for car (OScar) framework~\cite{rapelli2024oscar}, which provides a complete ETSI C-ITS protocol stack. The ArduSimple board enables \gls{gnss} positioning through \gls{rtk} services, achieving centimeter-level accuracy. Finally, Figure~\ref{fig:antenna-test} displays the MobileMark SMWG-313 antenna~\cite{SMWG-313}, a magnetic, automotive-grade, multi-band antenna designed to support both the \gls{gnss} receiver and two \gls{mimo} antenna elements for transmission and reception at \gls{v2x} frequencies.
An identical setup, as previously described, was deployed on two vehicles: one was parked at the roadside and served as the transmitter $T_x$, periodically sending vehicular \glspl{cam} at fixed time intervals; the other was in motion and acted as the receiver $R_x$. Two sets of field tests were conducted with $T_x$ and $R_x$---shown in Figure~\ref{fig:sali-sat} and Figure~\ref{fig:torino-sat}---during which a range of parameters was measured. Upon packet reception, $R_x$ recorded the payload, the UNIX timestamp of the event, the average measured \gls{rssi}, and its current GPS coordinates. 
The \glspl{dt} for ray tracing in VaN3Twin---Figure~\ref{fig:sali-sionna} and Figure~\ref{fig:torino-sionna}---were modeled in Blender using the Blosm add-on, which imports 2.5D representations of the propagation environment from OpenStreetMap. 
\firstrev{Following ITU-R P.2040 recommendations~\cite{iturp2040}, buildings were modeled as concrete structures. When present, trees were modeled as fully wooden objects, while the ground as concrete. Vehicles were included as full 3D meshes with metallic bodies to account for their scattering contribution. The ray tracing engine simulates specular reflection, diffuse reflection, and refraction, with a maximum of 5 interactions (bounces) per ray. Transmitting and receiving antennas were modeled as isotropic, which is a valid approximation of the antennas employed in the drive test.}
Simulation parameters used in the VaN3Twin ray tracer are summarized in Table~\ref{table:validazione-parameters}.
\begin{table}[!t]
    \caption{VaN3Twin parameters for ray tracing.}
    \label{table:validazione-parameters}
    \centering 
    \begin{tabular}{|p{15em} c|}
        \hline
        \multicolumn{2}{|c|}{\rule{0pt}{2.5ex}\rule[-0.9ex]{0pt}{2.5ex}\textbf{Ray tracing parameters}} \\
        \hline
        \rule{0pt}{2.5ex}\textbf{Carrier Frequency} [GHz] & 5.89 \\
        \rule{0pt}{2.5ex}\textbf{Buildings and Roads Material} & Concrete~\cite{iturp2040} \\
        \rule{0pt}{2.5ex}\textbf{Trees Material} & Wood~\cite{iturp2040} \\
        \rule{0pt}{2.5ex}\textbf{Cars Material} & Metal~\cite{iturp2040} \\
        \rule{0pt}{2.5ex}\textbf{Reflections} & \text{Specular, Diffused} \\
        \rule{0pt}{2.5ex}\textbf{Refractions} & \text{Yes} \\
        \rule{0pt}{2.5ex}\textbf{Maximum Interactions} & 5 \\
        \rule{0pt}{2.5ex}\textbf{Type of Antennas} & Isotropic \\
        \rule{0pt}{2.5ex}\textbf{Polarization} & Vertical \\
        \rule{0pt}{2.5ex}\rule[-0.8ex]{0pt}{2.5ex}\textbf{$T_x$ and $R_x$ Height} [m] & 1.8 \\
        \hline
    \end{tabular}
\end{table}
\firstrev{Although this work focuses on low-rate CAM traffic, VaN3Twin supports a wide range of application-layer profiles, including high-rate and latency-sensitive workloads such as cooperative perception and sensor sharing. CAM traffic was selected to enable direct validation against real-world measurements obtained using hardware limited to such traffic generation. Extending the evaluation to additional traffic profiles is straightforward, as the underlying ns-3 simulator supports arbitrary traffic patterns, including for example adaptive video streaming~\cite{liu2023simulationvideostreamingwireless}.}
Validation was carried out at two levels: (i) propagation, by comparing real $R_x$ RSSI measurements against their virtual counterparts; and (ii) packet reception, by assessing alignment between packet receptions in the virtual $R_x$ and the physical dataset. 
\firstrev{For (ii), we introduce a dedicated metric, the \gls{dr}, to assess the extent to which VaN3Twin enhances overall replication accuracy. The \gls{dr} quantifies the level of disagreement between VaN3Twin and a baseline in terms of binary decision outcomes. Let $\mathcal{R}$ and $\mathcal{M}$ denote the sets of decision outcomes, positive or negative, by VaN3Twin and by the baseline, respectively. The symmetric difference $\mathcal{R} \Delta \mathcal{M}$ then represents the set of outcomes classified as positive or negative by one, but not by the other, i.e., the instances where the two disagree on the decision. $\mathcal{R}\cup \mathcal{M}$ represents the totality of the decisions. The \gls{dr} is defined as:
\begin{equation}
    DR = \left|\mathcal{R}\Delta\mathcal{M}\right|/\left|\mathcal{R}\cup \mathcal{M}\right|.
\end{equation}
A lower value of \gls{dr} indicates a stronger alignment between the considered entities, whereas higher values highlight greater discrepancies in their decision outcomes.}
To enable a comprehensive evaluation of the differences between stochastic models (as adopted by current V2X state of the art simulation tools), a ray-based propagation model, and real-world measurements, the same simulation was also conducted using the vanilla version of ms-van3t.
In this configuration, ms-van3t relies on a stochastic propagation model based on 3GPP~TR~36.885~\cite{3gpp_tr_36_885} specifications, which incorporates distance-dependent, probabilistic \gls{los} status determination. More specifically, the \gls{los} probability is:
\begin{align} \label{eq:los-probability}
P_{LoS} = \min\left(\frac{18}{d}, 1\right)\left(1 - e^{-d/36}\right) + e^{-d/36},
\end{align}
where $d$ is the distance between the transmitter and the receiver. Unlike ray tracing, this model has no tunable parameters, as it is rigidly defined by 3GPP specifications. Notably, this model is well aligned with the conditions of both scenarios. The rural campaign featured a full \gls{los} between transmitter and receiver, closely matching theoretical assumptions. Meanwhile, in the urban scenario, the transmitter was placed at a height comparable to that of a typical cellular microcell, precisely the conditions for which the model was designed to operate. These represent ideal conditions for the model’s intended use, making the comparison particularly relevant. In conclusion, both the real-vehicle field tests and the simulations with VaN3Twin and ms-van3t used the IEEE 802.11p communication stack.

\subsection{LoS rural scenario}

This measurement campaign was carried out in a completely open and flat environment near Sali Vercellese, in the countryside surrounding Vercelli, Italy. The environment is free from obstacles such as buildings or trees. The transmitting vehicle traveled along a straight road, performing a round trip following the trajectory shown in Figure~\ref{fig:sali-sat} and Figure~\ref{fig:sali-sionna}. The moving vehicle, $R_x$, maintained an uninterrupted \gls{los} condition with the transmitting vehicle, $T_x$, which remained parked and stationary. This scenario serves as a necessary baseline for calibrating more complex experiments, such as those conducted in urban environments, where the presence of buildings and other obstacles significantly affects signal propagation.
\begin{figure} [!t]
    \centering
    \begin{subfigure}[t]{\linewidth}
        \centering
        \includegraphics[width=0.9\linewidth]{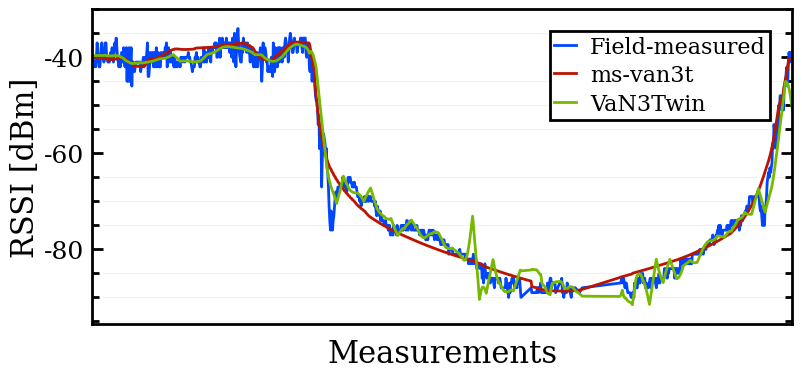}
        \caption{Comparison of the field-measured RSSI (blue) with the estimated value using VaN3Twin (green) and ms-van3t (red).}
        \label{fig:rssi-sali}
        \hfill
    \end{subfigure}
    \quad
    \begin{subfigure}[t]{\linewidth}
        \centering
        \includegraphics[width=0.9\linewidth]{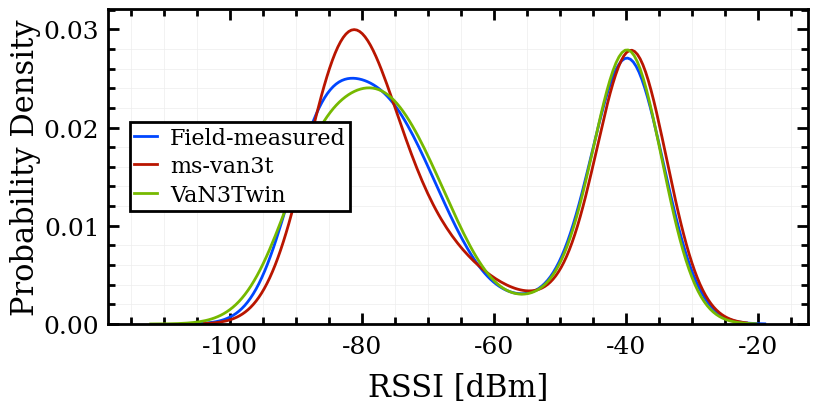}
        \caption{Probability Density Function of the field-measured RSSI (blue), the estimated value with VaN3Twin (green) and with ms-van3t (red).}
        \label{fig:pdf-sali}
        \hfill
    \end{subfigure}
    \caption{Measured and simulated RSSI values (a) and their Probability Density Function (b) for the rural measurement campaign in Sali Vercellese, Vercelli, Italy.}
    \label{fig:results-sali}
\end{figure}
Figure~\ref{fig:results-sali} presents a comparison of signal propagation across the different models. Specifically, the values obtained from field measurements are shown in blue, those generated by VaN3Twin in green, and the ones with ms-van3t in red. In Figure~\ref{fig:rssi-sali}, which depicts the \gls{rssi} values, it can be observed that the measured signal closely follows the trend reproduced by the VaN3Twin simulation, with consistently aligned values along the entire trajectory of the moving vehicle $R_x$. As expected, ms-van3t also aligns well with the measured data, given the simplicity of the scenario. However, it is worth noting that the \gls{rssi} trend produced by ms-van3t appears overly smooth and lacks local fluctuations, an inherent limitation of its theoretical formulation which does not account for small-scale propagation effects. This observation is further supported by Figure~\ref{fig:pdf-sali}, where the \glspl{pdf} of the three datasets exhibit overall consistent trends.

Regarding the Packet Reception \glspl{dr}, a disagreement of 12.82\%  between the real-world measurements and the simulation results is obtained using ms-van3t. Despite the simplicity of the propagation environment, the use of VaN3Twin nearly halves this discrepancy, reducing it to only 6.5\%.

\subsection{NLoS urban scenario}
Compared to the previously considered rural case, the urban scenario presents a significantly more complex propagation environment. These measurements were conducted in a particularly busy area near the city center of Turin, Italy, which is characterized by multiple trees, parked cars, and various other obstacles, causing a considerable increase in the number of interactions of the transmitted signal with its surroundings affecting the \gls{rssi}. Unlike the scenario in Sali Vercellese, the visibility conditions in this case vary depending on the relative positions of the vehicles within the environment. At the beginning of the measurement, the parked transmitting vehicle $T_x$ and the moving receiver $R_x$ are in a \gls{nlos} condition. As illustrated in Figure~\ref{fig:torino-sat} and Figure~\ref{fig:torino-sionna}, the transmitter follows a straight trajectory from right to left. Around the midpoint, a brief \gls{los} interval occurs before the vehicles return to a \gls{nlos} configuration, which persists until the end of the test. This pattern replicates a typical urban scenario, in which two vehicles located on perpendicular roads attempt to communicate with one another.
\begin{figure} [!b]
    \centering
    \begin{subfigure}[t]{\linewidth}
        \centering
        \includegraphics[width=0.9\linewidth]{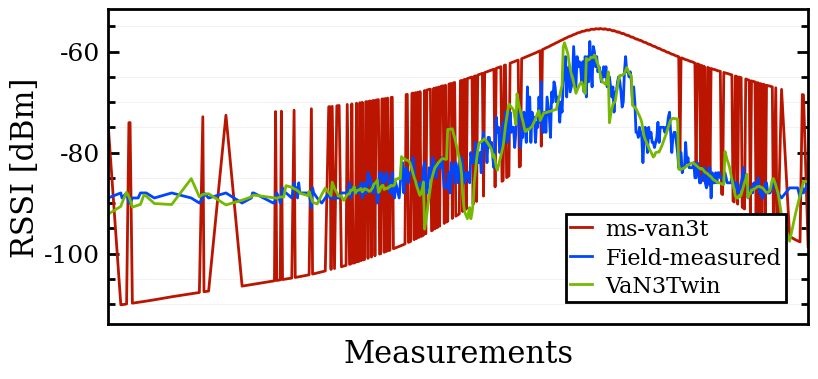}
        \caption{Comparison of the field-measured RSSI (blue) with the estimated value using VaN3Twin (green) and ms-van3t (red).}
        \label{fig:rssi-torino}
        \hfill
    \end{subfigure}
    \quad
    \begin{subfigure}[t]{\linewidth}
        \centering
        \includegraphics[width=0.9\linewidth]{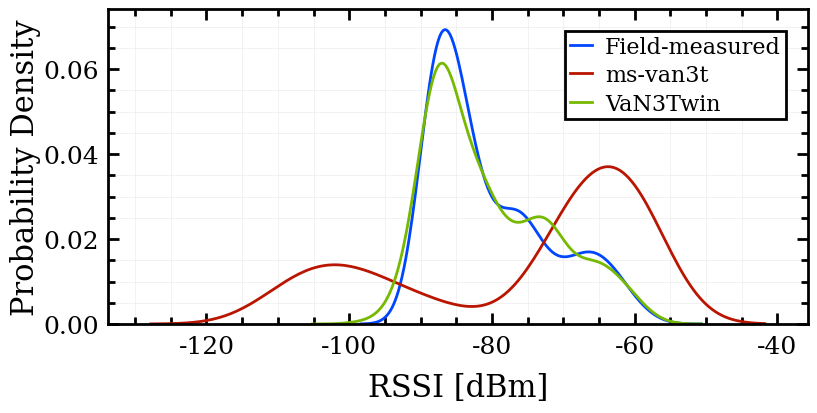}
        \caption{Probability Density Function of the field-measured RSSI (blue), the estimated value with VaN3Twin (green), and with ms-van3t (red).}
        \label{fig:pdf-torino}
        \hfill
    \end{subfigure}
    \caption{Measured and simulated RSSI values (a) and their Probability Density Function (b) for the urban measurement campaign in Turin, Italy.}
    \label{fig:results-torino}
\end{figure}
Similarly to Figure~\ref{fig:results-sali}, Figure~\ref{fig:results-torino} presents a comparison of signal propagation across the different models for the \gls{nlos} urban scenario. The results in Figure~\ref{fig:rssi-torino} show a reasonable alignment between the \gls{rssi} values obtained from field measurements and those extracted from VaN3Twin. Although the match is less precise than in the quasi-ideal rural case---due to the increased complexity of the urban environment---the overall trends remain consistent. It is important to note that while VaN3Twin offers a high-fidelity reproduction of realistic channel conditions, it cannot guarantee a perfect match in all situations. In this case, for instance, augmenting the level of detail of the virtual scene with additional elements such as parked vehicles, trees, and street furniture would likely reduce the residual gap and further improve simulation accuracy. Conversely, the output of vanilla ms-van3t reveals evident limitations in this context. The graph shows abrupt and continuous RSSI fluctuations, results of the rigid alternation between \gls{los} and \gls{nlos} propagation models as per equation (\ref{eq:los-probability}).
This behavior is further highlighted in Figure~\ref{fig:pdf-torino}, where the \gls{pdf} generated by VaN3Twin closely mirrors the measured distribution, accurately capturing both the mean and the variability of the RSSI values. In contrast, the vanilla ms-van3t model yields a distinctly bimodal distribution which leads to unrealistic clustering and fails to reflect the smooth variations observed during the field test.
The computed for Packet Reception \gls{dr} in the \gls{nlos} urban scenario reveals a substantial disagreement of 27.9\% between real-world data and the simulation performed using the ms-van3t model. Also in this case, VaN3Twin significantly reduces the discrepancy, achieving a \gls{dr} of just 8.16\%. This result highlights that, although the 3GPP-based stochastic model in \mbox{ms-van3t} is specifically calibrated for urban scenarios with \gls{nlos} conditions, VaN3Twin achieves an accuracy level that is more than three times higher.
\begin{table}[!t]
    \caption{Packet Reception DRs of ms-van3t and VaN3Twin with measured data.}
    \label{table:application-dr}
    \centering 
    \begin{tabular}{|l|c|c|c|}
        \hline
        \rule{0pt}{2.5ex}\rule[-0.9ex]{0pt}{2.5ex}\textbf{Type of Scenario} & \textbf{ms-van3t} & \textbf{VaN3Twin} & \textbf{Difference} \\
        \hline
        \rule{0pt}{2.5ex}\textbf{Rural} (Sali Vercellese) & 12.82\% & 6.5\% & 49.3\% \\
        \rule{0pt}{2.5ex}\rule[-0.9ex]{0pt}{2.5ex}\textbf{Urban} (City of Turin) & 27.9\% & 8.16\% & 70.8\% \\
        \hline
    \end{tabular}
\end{table}
In conclusion, given its deterministic and ray-based propagation model, VaN3Twin is capable of reproducing even complex, urban environments, leading to a higher punctual accuracy at the applicative level when compared to the stochastic model in vanilla ms-van3t. Table~\ref{table:application-dr} summarizes the computed Packet Reception \glspl{dr} using ms-van3t and VaN3Twin to reproduce real-world measurements, showing decreases of 49.3\% in the rural scenario and 70.8\% in the urban scenario. Building on this capability, the following section introduces the VaN3Twin coexistence module, which accounts for interference generated by other transmitting vehicles operating in parallel with different \glspl{rat}, thereby further increasing the realism and completeness of the simulation.

\section{VaN3Twin for multi-rat coexistence} \label{sec:coexistence-main-section}


Given the pivotal role of the ray tracing engine in VaN3Twin for reconstructing a consistent and realistic channel model, we leveraged this capability to develop a dedicated {\it coexistence module}. While the vanilla version of ms-van3t could simulate scenarios where vehicles using different \glspl{rat} coexist, it did not account for the interference that transmissions from one \gls{rat} might cause to others. To address this limitation, we designed a module to explicitly model inter-\gls{rat} interference, thereby enabling accurate evaluation of communication performance in multi-\gls{rat} scenarios.
This section introduces the underlying logic of the VaN3Twin module developed to manage multi-\gls{rat} coexistence, accounting for partial or full spectrum overlap between concurrent transmissions across different vehicular communication technologies. Finally, we compare three configurations: the vanilla ms-van3t model without coexistence support, the ms-van3t propagation model coupled with the coexistence module, and the coexistence module integrated with the accurate channel modeling of VaN3Twin.



\subsection{Proposed coexistence implementation} 
\label{sec:coexistence}

The vanilla version of ms-van3t includes different default propagation models for each \gls{rat}.
However, the LTE-V2X and NR-V2X modules typically rely on 3GPP-based propagation models \cite{3gpp_tr_36_885, 3gpp_tr_38_901} that are incompatible with those used by the IEEE 802.11p module, and vice versa. Furthermore, as previously discussed, the default channel models in ms-van3t employ probabilistic \gls{los} determination, limiting their accuracy in complex scenarios. To partially mitigate these limitations, ms-van3t allows manual adjustment of the channel condition through external inputs. This can be achieved either by directly injecting the \gls{los} status---e.g., from a ray tracing engine---or by generating a 2D building map within the simulator for automatic \gls{los}/\gls{nlos} classification. However, the latter approach is both time-consuming and imprecise, as it represents buildings using basic box geometries. In contrast, VaN3Twin leverages ray tracing to deliver significantly improved modeling accuracy and a unified channel abstraction that is consistent across different \glspl{rat}. This unified approach enables consistent and realistic channel estimation, as previously formalized in equation~\eqref{eq:channel_matrix_time}. Such consistency is especially critical for multi-\gls{rat} coexistence scenarios, where realistic modeling of shared or partially overlapping spectrum is required.
In a V2X simulation scenario, let $\mathcal{W}$ denote the set of simulated \glspl{rat} for which the coexistence module is enabled. Each $w \in \mathcal{W}$ operates over a frequency band of bandwidth $B^{(w)}$, centered at carrier frequency $f_c^{(w)}$, and utilizes an \gls{ofdm} waveform with $L^{(w)}$ subcarriers spaced by $\Delta f^{(w)}$. The corresponding sub-channel grid is given by: 
\begin{equation*}
    \mathcal{F}^{(w)} = \left\{ f_c^{(w)} + l \cdot \Delta f^{(w)} \right\}~\text{with}
\end{equation*}
\begin{equation}
    f_c^{(w)} + l \cdot \Delta f^{(w)} \in
    \left[ f_c^{(w)} - \frac{B^{(w)}}{2},~f_c^{(w)} + \frac{B^{(w)}}{2} \right],
\end{equation}

where $l \in L^{(w)}$ is an integer. To evaluate inter-\gls{rat} interference, a global resource grid must be defined that encompasses all technologies in the set $\mathcal{W}$. For each $w \in \mathcal{W}$, the corresponding grid is defined as:
\begin{equation}
    \mathcal{Z}^{(w)} = \mathcal{F}^{(w)} \times \mathcal{S}^{(w)},
\end{equation}
where $\mathcal{S}^{(w)}$ is the set of corresponding time slots.
The overall resource grid is defined as:
\begin{align}
    \mathcal{Z} = \mathcal{F} \times \mathcal{S},~
    \text{with}~~ 
    \mathcal{F} = \bigcup_{w \in \mathcal{W}} \mathcal{F}^{(w)},~~
    \mathcal{S} = \bigcup_{w \in \mathcal{W}} \mathcal{S}^{(w)}. 
\end{align}

Given a resource block $z \in \mathcal{Z}$, let $\mathcal{W}(z) \subseteq \mathcal{W}$ denote the set of \glspl{rat} actively using it. To monitor potential inter-\gls{rat} interference, the coexistence module continuously observes the communication channel and tracks all transmissions from each $w \in \mathcal{W}(z)$.\footnote{This feature enables concurrent simulation of IEEE 802.11p and NR-V2X.} To support this functionality, a set of tracking objects $\mathcal{H}$ is introduced. For each transmitting vehicle $v_n \in \mathcal{V}_{\text{tx}}(t) \subseteq \mathcal{V}_{\text{active}}(t)$ at time $t$, a corresponding tracking object $H_{v_n} \in \mathcal{H}$ is defined as:
\begin{align}
    H_{v_n} = \left( w \in \mathcal{W},~f_c^{(w)},~B^{(w)},~T,~S(f) \right),
\end{align}
where $T$ is the transmission duration and $S(f)$ the Power Spectral Density (PSD).
Let us now consider two vehicles transmitting using technologies $w$ and $w'$, both belonging to $\mathcal{W}(z)$, with $w \neq w'$. The SINR on resource block $z$ for the transmission using technology $w$ is computed as:
\begin{align}\label{eq:sinr-resblock}
    \text{SINR}_{w,z} = \frac{P_{\text{Rx}, w, z}}{N_{0,z} + P_{\text{int}, w} + \sum_{w' \neq w} P_{\text{Rx}, w', z}},
\end{align}
where $P_{\text{Rx}, w, z}$ denotes the received power of the desired signal, $N_{0,z}$ the noise power on $z$, $P_{\text{int}, w}$ the intra-RAT interference from other vehicles using $w$, and $P_{\text{Rx}, w', z}$ the interfering power from technology $w'$. The contribution of $P_{\text{Rx}, w', z}$ in equation~\eqref{eq:sinr-resblock} is computed as follows: let $v_{\text{int}}$ be an interfering vehicle using technology $w' \in \mathcal{W}(z)$, with an associated tracking object $H_{v_{\text{int}}}$. To discard negligible interference, a threshold on the total path gain $\mathcal{G}_{\text{thr}}$ is introduced. For a receiving vehicle $v_r$ using technology $w \neq w'$, interference from $v_{\text{int}}$ is ignored if the total path gain $\mathcal{G}_{v_{\text{int}}, v_r}$ is less than the minimum threshold, i.e., $\mathcal{G}_{v_{\text{int}}, v_r} < \mathcal{G}_{\text{thr}}$. Otherwise, the received interference power is calculated based on the PSD $S$ and the total path gain $\mathcal{G}_{v_{\text{int}}, v_r}$.
The overall SINR for technology $w$ is obtained by computing a bandwidth-weighted average over all resource blocks $z \in \mathcal{Z}$:
\begin{align}
    \text{SINR}_{w} = \frac{1}{|\mathcal{Z}|} \sum_{z \in \mathcal{Z}} \omega_z \cdot \text{SINR}_{w,z},
\end{align}
where $\omega_z$ is a weight related to the bandwidth of block $z$. 
\begin{algorithm} [!t]
\caption{SINR evaluation with inter-RAT contributions}
\begin{algorithmic}[1]
\REQUIRE Set of active \glspl{rat} $w \in \mathcal{W}$
\REQUIRE Sub-channels $\mathcal{F}^{(w)} ~ \forall ~ w \in \mathcal{W}$ 
\REQUIRE Time slots $\mathcal{S}^{(w)} ~ \forall ~w \in \mathcal{W}$

\FORALL{$w \in \mathcal{W}$}
    \STATE Set $\mathcal{F} \gets \mathcal{F} \cup \mathcal{F}^{(w)}$
    \STATE Set $\mathcal{S} \gets \mathcal{S} \cup \mathcal{S}^{(w)}$
\ENDFOR

\STATE $\mathcal{Z} \gets \mathcal{F} \times\mathcal{S} $
\STATE Track each $v_n \in \mathcal{V}_{\text{tx}}(t) \subseteq \mathcal{V}_\text{active}(t)$

\FORALL{blocks $z \in \mathcal{Z}$}
    \STATE Identify set of active \glspl{rat} $\mathcal{W}(z) \subseteq \mathcal{W}$
    
    \FORALL{RATs $w \in \mathcal{W}(z)$}
        \STATE Initialize interference to only intra-RAT contributions
        \STATE Set $P_{\text{int,tot}} \gets  P_{\text{int}, w} $
        \FORALL{$w' \in \mathcal{W}(z),~w' \neq w$}
            \FORALL{tracked interferers $v_\text{int}$ using $w'$}
                \IF{total path gain $\mathcal{G}_{v_{\text{int}}, v_r} \geq \mathcal{G}_{\text{thr}}$}
                    \STATE $P_{\text{int,tot}} \gets P_\text{int,tot} + P_{\text{Rx},w',z}$
                \ENDIF
            \ENDFOR
        \ENDFOR
        \STATE Compute $\text{SINR}_{w,z}$
    \ENDFOR
\ENDFOR

\FORALL{$w \in \mathcal{W}$}
    \STATE Compute average $\text{SINR}_{w}$ weighted by $\omega_z$
\ENDFOR

\end{algorithmic}
\label{algo:sinr}
\end{algorithm}
For clarity and ease of understanding, the SINR computation process described above is also summarized in Algorithm~\ref{algo:sinr}. 

Furthermore, to illustrate the capabilities of the coexistence module, a simulation of a vehicular scenario with multi-\glspl{rat} communication is presented and analyzed in the following subsection. 
The goal is not to conduct an exhaustive co-channel interference study for vehicular multi-\gls{rat} simulations, but to evaluate how SINR computation and its impact at the application level differ when interference from a coexisting technology is modeled using VaN3Twin versus ms-van3t. This comparison quantifies the benefits of integrating accurate physical-layer modeling within an \gls{ndt} framework.


\subsection{Multi-RAT simulation with coexistence module}
To validate the coexistence mechanism under realistic interference conditions, a multi-\glspl{rat} scenario is considered in which 20 vehicles traverse an urban environment containing buildings and obstacles, resulting in both \gls{los} and \gls{nlos} propagation conditions. \firstrev{The \gls{dt} was created following the same procedure introduced in Section~\ref{sec:validation} while the ray tracing parameters are the ones in Table~\ref{table:validazione-parameters}}. Each vehicle periodically transmits \glspl{cam} and \glspl{cpm} following ETSI transmission rules, using a common carrier frequency of 5.9~GHz and a bandwidth of 10~MHz.
The vehicles are evenly split between two technologies: 10 use IEEE 802.11p and 10 use NR-V2X.
\firstrev{Table~\ref{tab:parameters1} and Table~\ref{tab:parameters2} summarize the main simulation parameters for NR-V2X and IEEE 802.11p communication technologies, respectively. The setup is implemented in VaN3Twin with an ns-3 simulation configured to define the communication environment. The configuration process begins by reading the .xml route file from SUMO, which defines the vehicles to be generated. Next, the ns-3 nodes are instantiated, with each node representing a C-ITS station. Each node is equipped with the necessary Facilities, such as the Cooperative Awareness (CA) Basic Service, followed by the creation of the PHY and MAC layers for each communication technology in use. Once all components are properly initialized, the simulation starts, and synchronization between ns-3, SUMO, and Sionna RT proceeds as previously described.}
\begin{table}[!b]
    \caption{VaN3Twin parameters for NR-V2X.\label{tab:parameters1}}
    \centering
    \begin{tabular}{|lc|}
    \hline
    \multicolumn{2}{|c|}{\rule{0pt}{2.5ex}\rule[-0.9ex]{0pt}{2.5ex}\textbf{NR-V2X Parameters}} \\
    \hline
    \rule{0pt}{2.5ex}\textbf{Carrier Frequency} [GHz]   & 5.9 \\
    \rule{0pt}{2.5ex}\textbf{Channel Bandwidth} [MHz]   & 10 \\
    \rule{0pt}{2.5ex}\textbf{Subchannel Bandwidth} [MHz] & 10 \\
    \rule{0pt}{2.5ex}\textbf{Numerology}                & 2 \\
    \rule{0pt}{2.5ex}\textbf{Modulation and Coding Scheme} & 14 \\
    \rule{0pt}{2.5ex}\textbf{Sidelink Threshold For PSSCH (RSRP)} [dB]     & 21 \\
    \rule{0pt}{2.5ex}\textbf{Resource Blocks For PSSCH}   & 10 \\
    \rule{0pt}{2.5ex}\textbf{Number Of TX Per Resource Block}         & 3 \\
    \rule{0pt}{2.5ex}\textbf{Reservation Period} [ms]   & 20 \\
    \rule{0pt}{2.5ex}\textbf{PHY Modulation}                  &  OFDM  \\
    \rule{0pt}{2.5ex}\textbf{Subcarrier Spacing} [kHz]  & 60  \\
    \rule{0pt}{2.5ex}\textbf{Tx Power} [dBm]                  & 30 \\
    \rule{0pt}{2.5ex}\textbf{Sensitivity} [dBm]         & -93 \\
    \rule{0pt}{2.5ex}\textbf{SINR Threshold} [dB]       & 10 \\
    \rule{0pt}{2.5ex}\textbf{Resource Selection Method}              & Random~\cite{sandra_lagen} \\
    \rule{0pt}{2.5ex}\textbf{Propagation Loss Model}   & 3GPP-based~\cite{3gpp_tr_38_901} \\
    \rule{0pt}{2.5ex}\textbf{Sidelink Mode}   & PC5 Mode 4 \\
    \hline
    \end{tabular}
\end{table}
\begin{table}[!b]
    \caption{VaN3Twin parameters for IEEE 802.11p.}
    \label{tab:parameters2}
    \centering
    \begin{tabular}{|lc|}
    \hline
    \multicolumn{2}{|c|}{\rule{0pt}{2.5ex}\rule[-0.9ex]{0pt}{2.5ex}\textbf{IEEE 802.11p Parameters}} \\
    \hline
    \rule{0pt}{2.5ex}\textbf{Carrier Frequency} [GHz]   & 5.9 \\
    \rule{0pt}{2.5ex}\textbf{Channel Bandwidth} [MHz]   & 10 \\
    \rule{0pt}{2.5ex}\textbf{PHY Modulation}                  &  OFDM  \\
    \rule{0pt}{2.5ex}\textbf{Tx Power} [dBm]                  & 30 \\
    \rule{0pt}{2.5ex}\textbf{Sensitivity} [dBm]         & -93 \\
    \rule{0pt}{2.5ex}\textbf{SINR Threshold} [dB]       & 10 \\
    \rule{0pt}{2.5ex}\textbf{Data Rate} [Mb/s]          & 3 \\
    \rule{0pt}{2.5ex}\textbf{Carrier Access Protocol}              & CSMA/CA \\
    \rule{0pt}{2.5ex}\textbf{Propagation Loss Model}    & 3GPP-based~\cite{3gpp_tr_36_885} \\
    \hline
    \end{tabular}
\end{table}
Since all vehicles operate on the same frequency band, transmissions from one technology can act as interference for the other, depending on spectral overlap and propagation conditions.
In order to establish a baseline representative of the current state of the art, the simulation is first conducted in ms-van3t without the coexistence module. Subsequently, the same simulation is repeated with the coexistence module enabled, using both ms-van3t and VaN3Twin. This procedure allows for the assessment of how the different propagation modeling approaches---3GPP-based in ms-van3t and ray-based in VaN3Twin---affect the estimated SINR for each received packet, the identification of potential interferers, and ultimately the outcome of the packet decoding process. To ensure identical propagation conditions for both models, \gls{los} determination in \mbox{ms-van3t} was injected using information from VaN3Twin’s ray tracer, enabling a fair and meaningful comparison.
\begin{figure}[!b]
    \centering
    \includegraphics[width=0.9\linewidth]{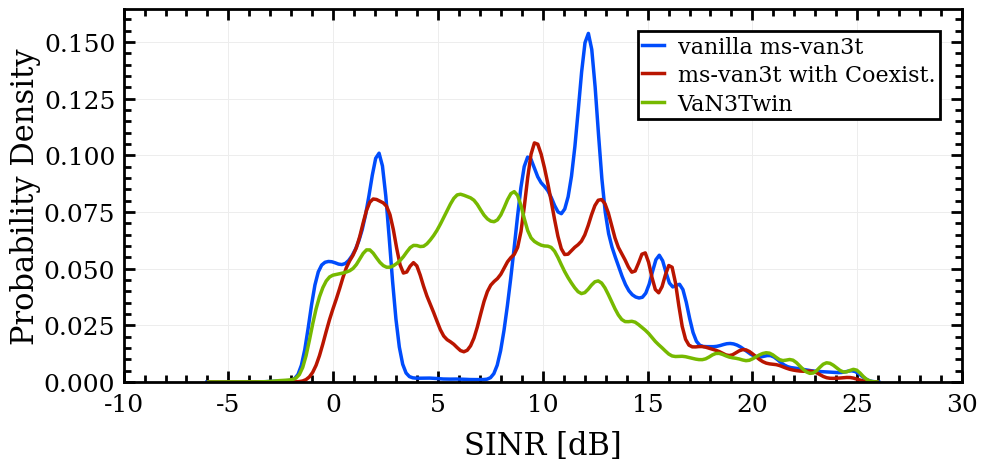}
    \caption{Probability Density Functions of the measured SINRs for all the simulated V2V links computed using vanilla ms-van3t without coexistence module (blue), ms-van3t with coexistence module (red), and VaN3Twin (green).}
    \label{fig:generic_pdf}
\end{figure}
Figure~\ref{fig:generic_pdf} shows the SINR \glspl{pdf} produced by all three considered simulators. While the distribution obtained using VaN3Twin exhibits a Gaussian-like trend, the ones generated by the two versions of ms-van3t both display two prominent peaks. This bimodal characteristic stems from the rigid model switching governed by the binary \gls{los}/\gls{nlos} classification of the communication link, as already observed during the validation phase---see Figure~\ref{fig:results-torino}. In real-world propagation environments, however, the transition between \gls{los} and \gls{nlos} conditions is not discrete but rather gradual. Even under \gls{los} conditions, surrounding obstacles can cause signal refraction and attenuation, leading to a degradation in reception quality. Such effects are not captured by the discrete switching logic of ms-van3t, but are instead considered in the ray tracer behavior of VaN3Twin. More generally, Figure~\ref{fig:generic_pdf} highlights a tendency of simulators using 3GPP-based propagation models to overestimate SINR values, even if the deterministic \gls{los} state information from the ray tracing engine is injected. The absence of coexistence-aware modeling in the vanilla ms-van3t configuration leads to significantly higher SINR estimates.

Another critical aspect in vehicular communication systems is the impact of surrounding vehicles, which can give rise to \gls{nlosv} conditions. These scenarios cannot be neglected in realistic modeling. For example, millicar~\cite{drago2020millicar}, a V2V module for ns-3 specifically designed for FR2 communications, addresses this issue through a probabilistic approach that models both the likelihood of vehicular obstructions and the statistical distribution of their physical dimensions. As discussed in Section~\ref{subsec:greenflow}, the ray tracing engine provides a binary output $\lambda_{v_i,v_j}$ indicating the \gls{los} status between two vehicles $(v_i, v_j)$. This determination is based solely on the presence or absence of at least one \gls{los} path within the set of admissible propagation rays $\mathcal{R}_{v_i,v_j}$. 
Moreover, the currently available models in ms-van3t do not support \gls{nlosv} conditions. Consequently, such cases are conservatively treated in ms-van3t as \gls{nlos} due to large obstacles as buildings.
\begin{figure}[!t]
    \centering
    \includegraphics[width=0.9\linewidth]{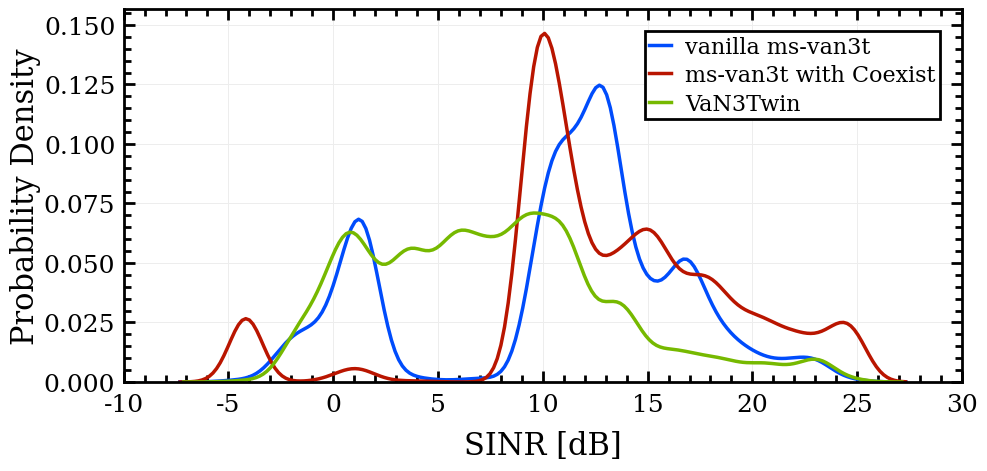}
    \caption{Probability Density Functions of SINRs for V2V links with frequent NLoSv scenarios computed using vanilla ms-van3t without coexistence module (blue), ms-van3t with coexistence module (red), and VaN3Twin (green).}
    \label{fig:nlosv_pdf}
\end{figure}
This limitation is evident in Figure~\ref{fig:nlosv_pdf}, which presents the SINR \gls{pdf} for a subset of vehicles frequently experiencing \gls{nlosv} conditions during the simulation. Here, the distribution obtained with VaN3Twin retains the previously observed Gaussian-like profile. Conversely, the two versions of ms-van3t notably overestimate SINR---irrespective of whether \gls{nlosv} conditions are treated as \gls{nlos}---both showing a pronounced peak around 10~dB.
In conclusion, the Gaussian-like distributions observed in the VaN3Twin results of Figures~\ref{fig:generic_pdf} and \ref{fig:nlosv_pdf} reflect a more realistic propagation behavior, as discussed in Section~\ref{sec:validation}.
As detailed in Section~\ref{sec:coexistence}, identifying nearby transmitters using other \glspl{rat} as potential sources of interference directly shapes the resulting SINR. This estimation, ultimately, determines whether an ongoing transmission is decodable at the receive or not. The combination of these mechanisms allows \mbox{VaN3Twin} to form a decision about the expected reception outcome, which may differ from that of \mbox{ms-van3t}.
\begin{figure} [t]
    \centering
    \includegraphics[width=0.9\linewidth]{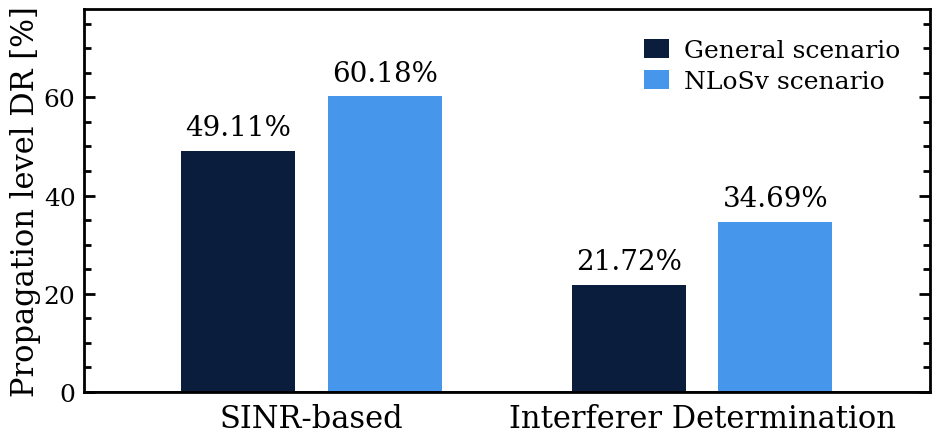}
    \caption{Propagation level Disagreement Ratios in interferer detection and SINR-based packet decodability decisions for a general scenario and for the ones with frequent \gls{nlosv}.}
    \label{fig:physical-dr}
\end{figure}
Figure~\ref{fig:physical-dr} reports two Propagation level \glspl{dr} for the two previously considered sets of simulated vehicles. The \gls{dr} for interferer detection between \mbox{ms-van3t} with coexistence enabled and \mbox{VaN3Twin} is 21.72\%, reaching 34.69\% in scenarios with frequent \gls{nlosv} conditions. This indicates that in up to 35\% of the cases, the \mbox{ms-van3t} incorrectly discards an interfering signal—treating it as negligible---whereas \mbox{VaN3Twin} correctly identifies it as having a significant impact on the transmission. These incorrect assumptions, combined with the higher modeling accuracy of VaN3Twin, also propagate to the final decision on packet decodability, further increasing the divergence between the two models. In the general case, the \gls{dr} reaches 49.11\%, meaning that \mbox{ms-van3t} and VaN3Twin disagree on packet reception in nearly half of the instances. This disagreement rises to 60.18\% under frequent \gls{nlosv} conditions, highlighting how \mbox{ms-van3t} is unable to capture transient obstructions.
\begin{figure}[!b]
    \centering
    \includegraphics[width=1\linewidth]{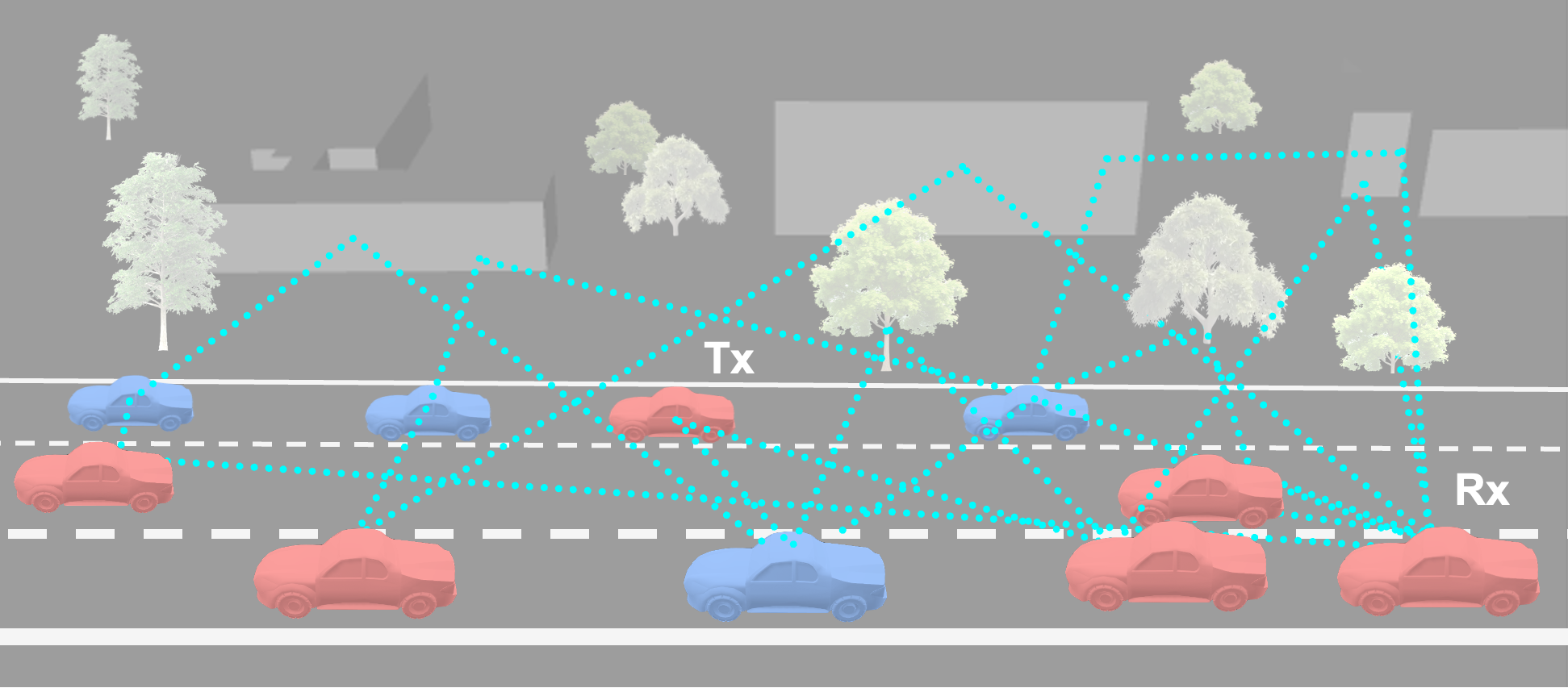}
   \caption{Simulation snapshot of an NR-V2X communication between Rx and Tx (red) interfered by 802.11p vehicles (blue) in a \gls{nlosv}-prone environment. Selected propagation paths from interferers to the receiver Rx are shown in blue.}
    \label{fig:nlosv-case-map}
\end{figure}
For example, for the communication between two NR-V2X vehicles shown in Figure~\ref{fig:nlosv-case-map}, ms-van3t and VaN3Twin yield differences in the estimated interfering power of up to 17.53~dBm, resulting in a corresponding SINR deviation of 4~dB.
As for the application layer performance, let us first consider the impact of the coexistence module on the overall \gls{prr} for the NR-V2X communication, which measures the final number of successfully processed packets. Using the vanilla version of ms-van3t, the resulting PRR is 97\%, decreasing to 93\% if the coexistence module is used. On the other hand, VaN3Twin computes a lower \gls{prr} of 84\%. The overall impact of inter-RAT interference on the \gls{prr} is then limited, primarily due to the small payload sizes of \glspl{cam} and \glspl{cpm} and the robust retransmission mechanisms provided by NR-V2X.
However, such aggregated indicators fail to capture discrepancies at the single packet level, crucial in a \gls{ndt}. Let us consider the Packet Reception \gls{dr} between ms-van3t with coexistence enabled and VaN3Twin. In the general case, we observe a \gls{dr} of 30.56\%, revealing a substantial mismatch between the two simulators. In the \gls{los}-dominant scenario with frequent \gls{nlosv} conditions, the \gls{dr} is slightly lower at 18.43\%. This reduction is primarily attributed to the higher likelihood of successful packet delivery at 5.9~GHz when vehicles are in close proximity, as is typical in such scenarios.

In conclusion, the overall impact of coexistence on aggregated packet delivery metrics appears limited. However, this apparent agreement at the application layer is somewhat misleading, as high packet reception \glspl{dr} were observed despite underlying modeling inaccuracies. This undermines the suitability of ms-van3t and, in general, of ns-3 as a \gls{ndt} component, which requires not only statistically accurate outcomes but also a faithful reproduction of real-world communication dynamics. Furthermore, significant discrepancies were observed in SINR estimation, especially during transient obstructions. These differences are expected to have a greater impact in more demanding use cases, where larger data payloads and higher spectral efficiency requirements intensify the need for accurate channel modeling, particularly at higher frequencies. Consequently, in scenarios involving advanced communication tasks---e.g., shared vision, cooperative perception, or platooning---the divergence between the two models is likely to become more pronounced. These use cases inherently demand higher SINR levels to ensure lower \gls{bler} values and to maximize the achievable transmission rate, thereby further exposing the limitations of ms-van3t under realistic vehicular conditions.

\subsection{Runtime and scalability analysis} 
\label{subsec:runtime}

\begin{table}[!t]
    \caption{\firstrev{VaN3Twin scalability analysis.}}
    \label{tab:scalability}
    \centering
    \begin{tabular}{|l|c|c|c|c|}
        \hline
        \rule{0pt}{2.5ex}\rule[-0.9ex]{0pt}{2.5ex}\textbf{Vehicles} & \textbf{Cache Hit} & \textbf{Speedup} & \textbf{Mean Cost} [ms] & \textbf{P95} [ms] \\
        \hline
        \rule{0pt}{2.5ex}\textbf{5}  & 92.65\% & 13.5$\times$  & 168.56 & 204.14 \\
        \rule{0pt}{2.5ex}\textbf{10} & 95.97\% & 24.7$\times$  & 196.30 & 253.25 \\
        \rule{0pt}{2.5ex}\textbf{20} & 98.46\% & 64.1$\times$  & 232.98 & 295.65 \\
        \rule{0pt}{2.5ex}\textbf{40} & 99.02\% & 150.6$\times$ & 252.03 & 468.66 \\
        \rule{0pt}{2.5ex}\textbf{60} & 99.58\% & 237.1$\times$ & 446.25 & 847.61 \\
        \hline
    \end{tabular}
\end{table}

\firstrev{The practical deployment of VaN3Twin on large-scale scenarios depends critically on the runtime cost of ray tracing computations. To quantify scalability, we recorded the cost---in terms of wall-clock time required to perform each channel parameter computation---with 5, 10, 20, 40, and 60 active vehicles, the maximum value allowed by the considered setup. Experiments were conducted on a workstation running Ubuntu 22.04.5 equipped with an NVIDIA GeForce RTX 3090 GPU with 25~GB of VRAM, 16 Intel Core i9-10980XE CPUs at 3.00~GHz, and 256~GB of RAM. The same scenario considered in Section~IV-B was used with ray tracing parameters for Sionna RT v1.2.0 as per Table~\ref{table:validazione-parameters} except for the maximum number of interactions, which is set to 3. VaN3Twin propagation paths caching subsystem presented in Section~II-C was enabled across all runs. The collected wall-clock times exhibit a strong bimodality introduced by the caching mechanism and therefore, in order to avoid biases introduced by the dominant reuse of cached propagation paths, we evaluate VaN3Twin scalability by jointly considering (i) the computational cost of full ray tracing computations between every active vehicle and (ii) the overall cache hit ratio. Observed statistics are summarized in Table~\ref{tab:scalability}. 
As the number of vehicles increases, two effects emerge. 
First, the mean cost of a full ray tracing snapshot increases predictably with the number of active vehicles, rising from 168.56 ms (5 vehicles) to 446.25 ms (60 vehicles), with the 95th percentile increasing from 204.14 ms to 847.61 ms. This behavior reflects the growth in scene complexity but remains limited thanks to the grouped computation of channels between active vehicle pairs and the GPU parallelization capabilities of Sionna RT. 
Second, the cache-hit ratio increases monotonically, reaching 99.58\% with 60 vehicles. In dense scenarios, each ray tracing snapshot produces channel information for a larger number of vehicle pairs, rapidly populating the cache and yielding a higher proportion of subsequent (almost) zero-cost operations, even though cache invalidation occurs more frequently. A critical factor is the overall system-wide acceleration,} \secondrev{with the joint application of VaN3Twin’s tightly coupled grouped channel computations and propagation path caching} \firstrev{yielding a 237$\times$ speedup relative to the baseline. As full ray tracing executions become comparatively rare at higher densities, their contribution to the amortized per-request cost is substantially reduced. This observation also suggests that increasing the age of the twin---i.e., balancing output accuracy with reduced update frequency from the physical network---combined with trajectory prediction to proactively schedule ray tracing snapshot computations, can further approach real-time operation. Within the tested range, VaN3Twin thus exhibits sublinear growth in amortized runtime and favorable reuse dynamics. These results indicate that high-fidelity ray tracing in large-scale V2X scenarios is computationally attainable, provided that propagation path reuse mechanisms are appropriately designed and dimensioned,}
\secondrev{together with the possibility to deploy such system as a federation of smaller, edge-located \glspl{ndt} operating over limited spatial areas.}

\section{Literature Review}

\begin{table*}[t]
\caption{\firstrev{Comparison of V2X frameworks with VaN3Twin, focusing on coexistence and ray tracing integration.}}
\label{table:related}
\begin{center}
\scriptsize
\begin{tabular}{ m{9.0em} >{\raggedright}m{13.5em} m{12.0em} m{11.0em} m{12.5em} m{5.0em} } 
 \hline
 \small \textbf{Framework/study} & \small \textbf{Supported RATs} & \small \vspace{2pt}\textbf{Coexistence\newline} \footnotesize{DSRC and C-V2X} & \small \textbf{Ray tracing} & \small \vspace{2pt}\textbf{V2X Protocols\newline} \footnotesize{Full-stack implementation} & \small\vspace{2pt}\textbf{Open source} \\
 \hline
 \textbf{WiLabV2XSim} \cite{co-exist-rel-1-todisco} & \vspace{3pt}IEEE 802.11p, LTE-V2X, NR-V2X & \checkmark & \textit{X} & \textit{X} & \checkmark \\ \\
 \textbf{LTEV2VSim} \cite{ltev2vsim} & IEEE 802.11p, LTE-V2X & \textit{X} & \textit{X} & \textit{X} & \checkmark \\ \\
 \textbf{Jayaweera et al.} \cite{co-exist-rel-4-jayaweera} & IEEE 802.11p, LTE-V2X & \textit{$\sim$} (MATLAB-based \newline numerical results) & \textit{X} & \textit{X} & \textit{X} \\ \\
 \textbf{Veins} \cite{veins-paper} & IEEE 802.11p, LTE-V2X \cite{opencv2x-mode4-model} & \textit{X} & \textit{X} (can be potentially\newline integrated as in \cite{RUZNIETO2023100964})  & \checkmark (ETSI stack requires\newline Artery \cite{artery_paper}) & \checkmark \\ \\
 \textbf{Rusca et al. }\cite{matlab-rt-colosseum} & Access technology-agnostic single RAT & \textit{X} & \checkmark (MATLAB RT) & \textit{X} & \textit{\checkmark} \\ \\
 \textbf{Gaugel et al.} \cite{gaugel2012accurate} & Access technology-agnostic single RAT & \textit{X} & \textit{$\sim$} (offline RT traces) & \textit{X} & \textit{X} \\ \\
 \textbf{ns3sionna} \cite{ns3-sionna-falko} & IEEE 802.11 & \textit{X} & \checkmark (Sionna RT in the loop) & \textit{X} & \checkmark \\ \\
 \textbf{DT-CoVeSS} \cite{Twardokus2505:DT-CoVeSS} & C-V2X & \textit{X} & \checkmark (customized Sionna RT) & \textit{$\sim$} (support for SAE J3161 and IEEE 1609.2) & \checkmark \\ \\
 \textbf{ms-van3t} \cite{ms-van3t-conference,ms-van3t-journal-2024} & IEEE 802.11p, LTE,\newline LTE-V2X, NR-V2X & \textit{X} & \textit{X} & \checkmark (ETSI stack with support to multiple versions \cite{ms-van3t-journal-2024}) & \textbf{\checkmark} \\ \\
 \textbf{VaN3Twin} & \textbf{IEEE 802.11p, LTE, LTE-V2X, NR-V2X\vspace{3pt}} & \textbf{\checkmark} & \textbf{\checkmark (Sionna RT in the loop)} & \textbf{\checkmark (ETSI stack with support to multiple versions \cite{ms-van3t-journal-2024})\vspace{3pt}} & \textbf{\checkmark} \\
 \hline
\end{tabular}
\end{center}
\end{table*}

This section reviews the state of the art in accurate V2X simulations and \glspl{ndt}, with particular emphasis on co-channel coexistence and the integration of ray tracing. We begin by analyzing prior work on coexistence studies and the extent to which open source simulators support them. We then examine the use of ray tracing as a key enabler for PHY-layer-accurate simulations in complex urban environments. Finally, we highlight the novelty of our approach compared to existing open source solutions\firstrev{, besides providing an overview on vehicular \glspl{dt}, to which VaN3Twin belongs.
Table~\ref{table:related} summarizes the differences between VaN3Twin and the main frameworks for V2X simulation and emulation currently available}.

\subsection{Simulation of co-channel coexistence}

\firstrev{Vehicular communications are a cornerstone for higher levels of driving automation, particularly at SAE levels 4 and 5~\cite{sae_levels_book_chapter}.}
\firstrev{In this context, }simulation plays a crucial role in the \gls{ndt} development, enabling the testing of new services in a safe and cost-effective manner~\cite{roongpraiwan2025digital, CazzellaV2XDT}. To this end, accurate simulation of the PHY and MAC layers, especially their interaction, is essential. While various studies have assessed each technology independently, both through simulation~\cite{ms-van3t-journal-2024,simulations-related-1,simulations-related-2} and field testing~\cite{field-tests-related-1,field-tests-related-2}, research on the coexistence of multiple technologies operating on the same channel remains scarce. 
For example, the study in~\cite{10424418} analyzes the interference that next-generation 6G networks may cause to passive sensing satellite systems.
As for V2X communications, studies like Maglogiannis et al.~\cite{co-exist-rel-3-maglogiannis}, consider separate channels, thereby excluding interference effects.
In~\cite{10884797}, Elloumi et al. propose a learning-based resource allocation framework for spectrum sharing in \gls{v2i} communications. However, their work does not implement a full-stack approach, focusing instead on interference-aware scheduling within a single \gls{rat}.
In particular, the coexistence of IEEE 802.11p and C-V2X Mode 4~\cite{tutorial-5g-nr-v2x} remains an open challenge. Both technologies share the congested 5.8–5.9~GHz spectrum, and their simultaneous deployment can degrade performance due to mutual interference~\cite{bazzi-co-existance-1}. 
In~\cite{bazzi-co-existance-1}, a possible solution is proposed involving an enhanced preamble for C-V2X, tested via the WiLabV2XSim simulator~\cite{co-exist-rel-1-todisco}, which is based on LTEV2VSim~\cite{ltev2vsim}. This MATLAB-based tool supports both IEEE 802.11p and C-V2X and has also been used in~\cite{co-exist-rel-2-manshaei} to assess different channel allocation strategies.
Recognizing the importance of coexistence, Jayaweera et al.~\cite{co-exist-rel-4-jayaweera} propose an optimization-based approach for dynamic technology and channel selection, though their simulations rely on MATLAB and do not involve a full-featured V2X stack.
While these tools can model different technologies, standard versions of simulators like ns-3 do not support mutual interference between Wi-Fi and cellular-based protocols\firstrev{\cite{ns-3-book-chapter}}. Extensions for C-V2X~\cite{eckermann-lte-v2x-model,zoraze-nr-v2x-model} and native IEEE 802.11p support exist, but precise coexistence modeling remains challenging. The ms-van3t framework, based on ns-3, SUMO~\cite{SUMO2018}, and CARLA~\cite{dosovitskiy2017carla}, supports multiple technologies and ETSI basic services, but prior to this work, detailed modeling of mutual interference was lacking.
OMNeT++~\cite{10.5555/1416222.1416290}, another popular network simulator, has been extended for V2X via frameworks like Veins~\cite{veins-paper} and OpenCV2X~\cite{opencv2x-mode4-model}. However, it does not offer an integrated approach to simulate coexistence between cellular and Wi-Fi-based technologies and presents a more complex architecture for real-world deployment compared to ns-3.
Before our work, WiLabV2XSim~\cite{co-exist-rel-1-todisco} was the only open source simulator capable of naively modeling co-channel interference between multiple V2X technologies.

\subsection{Integration of V2X simulators with ray tracers}
While simulators like ns-3, OMNeT++, and MATLAB-based tools differ in coexistence modeling, they commonly rely on stochastic or analytical channel models---e.g., Winner~\cite{winner_models}---which provide averaged metrics but struggle with accuracy in complex environments. These models often fail to capture interference at the packet level or the impact of environmental obstacles causing reflection, diffraction, and shadowing.
To overcome these limitations, integrating ray tracing into system-level vehicular simulations has been explored. For instance, Ruz Nieto et al.~\cite{RUZNIETO2023100964} integrated the Opal ray tracer~\cite{opal-ray-tracer}---\firstrev{an open source ray tracing engine based on NVIDIA OptiX, using a \gls{sbr} method}---with OMNeT++ to simulate LoRaWAN communications. While promising, the study does not include mobility or standardized V2X protocols.
In~\cite{matlab-rt-colosseum}, Rusca et al. developed a MATLAB-based RT framework using real mobility data to compute precise channel parameters. \firstrev{More specifically, the framework computes parameters such as path gain, phase shift, and angle of arrival, focusing on the PHY layer} and supporting \firstrev{advanced wireless channel emulation}, but lacks a standard-compliant V2X stack and requires commercial MATLAB licenses—an issue common to all MATLAB-based simulators using the built-in RT engine~\cite{matlab-rt}.
Gaugel et al.~\cite{gaugel2012accurate} proposed a more precise PHY model in ns-3 by incorporating ray tracer output, enabling bit-level simulation of packet reception. However, the integration is offline and does not support realistic mobility or a complete vehicular stack.
A more integrated solution, \firstrev{named \textit{ns3sionna}}, was recently presented by Zubow et al.~\cite{ns3-sionna-falko}, who coupled Sionna RT with ns-3 to simulate IEEE 802.11ac. Their approach relies on ZeroMQ for inter-process communication and focuses on indoor and outdoor WLAN scenarios.
In contrast, our work integrates Sionna RT into ms-van3t using lightweight UDP sockets for low-latency IPC \firstrev{as introduced by some of the authors of this paper in \cite{Pegu2505:Toward}}. This approach enables accurate simulations that support not only PHY and MAC layers but the entire V2X stack, including support for multiple \glspl{rat} and realistic vehicle mobility via SUMO, CARLA or real GPS coordinates gathered from the physical world. The proposed framework is uniquely suited for high-fidelity studies of multiple technologies and additionally enables their simultaneous simulation, supporting coexistence analyses in dynamic, multi-technology vehicular environments. 
Another example specifically targeting V2X communications was recently presented by Twardokus et al.~\cite{Twardokus2505:DT-CoVeSS}, who conducted a security evaluation of \gls{cv2x} leveraging a high-fidelity \gls{dt} using Sionna RT. However, their work does not account for the coexistence of \gls{dsrc} vehicles operating alongside \gls{cv2x} in the same environment.

\subsection{Comparison between existing ray tracers}
\firstrev{The computational efficiency of the selected ray tracing engine is critical for large-scale wireless simulations, especially in dynamic vehicular scenarios. Among the methods for ray tracing performance evaluation, benchmarking---i.e., measuring wall-clock execution time on representative inputs---offers a practical alternative to theoretical complexity analysis, as it does not require access to the internal algorithmic structure, not available for closed source commercial software.
Following this empirical approach, Zhu et al.~\cite{zhu2024toward} compared three ray tracing engines: Remcom Wireless InSite~\cite{Wireless-InSite}, Sionna RT~\cite{sionna}, and MathWorks' Ray Tracing Toolbox~\cite{matlab-rt}. Tests were performed on two scenarios: a simplified urban layout with low-resolution meshes, and a photorealistic digital twin generated via CARLA with complex, high-resolution geometry. Results show that Sionna RT consistently achieved the lowest execution times across both scenarios, with especially strong performance in complex scenes when configured to leverage on GPU acceleration and parallelization. Its runtime also scaled well with scene complexity, highlighting its robustness for realistic simulations. However, speed alone does not guarantee accuracy. Therefore, the adoption of Sionna RT within VaN3Twin is not only based on performance alone, but also on the validation of the overall framework against real-world measurements (see Section~\ref{sec:validation}). Furthermore, Sionna RT is the only open source option evaluated, which makes it preferable in terms of reproducibility and extensibility.}

\subsection{Vehicular digital twins}

\firstrev{In recent years, the concept of Digital Twin has gained significant attention as a way to define platforms able to model realistic scenarios and analyze their evolutions through simulated processes. While \glspl{dt} can be created for many physical systems---e.g., a \gls{dt} of the mechanical part of a vehicle---they are increasingly adopted to model the behaviors of networks, including vehicular ones with their related mobility patterns. In this context, real-world data are continuously provided to the the V2X \gls{dt} allowing its models to process them and make decisions subsequently fed back in closed-loop to the real vehicular network. The V2X \gls{dt} can also be employed to model in advance real-world scenarios---either from environment and mobility models or from previously collected data---to study system evolution under different configurations, for instance to support pre-deployment evaluations of new technologies. Moreover, hybrid approaches are possible, where the V2X \gls{dt} emulates additional entities that communicate in real time with real-world components, effectively enabling virtual hardware integration while reducing experimental costs. This specific capability is also supported by the VaN3Twin native emulation mode.}
\firstrev{For a V2X \gls{dt} to operate in the loop, parallel processing is often required to ensure near-real-time performance. A representative example is the RAVEN~\cite{RAVEN-paper} platform, which combines vehicle data ingestion with contextual information to predict vehicle positions and channel conditions, thereby managing interference and maximizing the capacity of multi-hop mmWave vehicular networks. Another notable case is the Vehicle-to-Cloud ADAS framework~\cite{V2C-ADAS}, where a cloud-based \gls{dt} performs ADAS decision processes that are subsequently used as actuation guidance. Additional V2X DT frameworks include TuST~\cite{TuST-paper-conference, TuST-paper-journal}, a large-scale mobility model replicating the entire city of Turin, Italy, from processed real traffic data. Similarly, Rusca et al.\cite{matlab-rt-colosseum} employ recorded mobility traces to construct accurate digital representations of vehicular environments for wireless emulation, while\cite{9861008} presents a \gls{dt} of a metro station reproducing the mobility dynamics of a complete indoor pedestrian scenario.}

\section{Conclusion}
This work introduces VaN3Twin\footnote{VaN3Twin \secondrev{and the simulation configuration files and their ray tracing scenarios used for gathering the results presented in this article} are available at this GitHub repository: \url{https://github.com/DriveX-devs/VaN3Twin}}, a \firstrev{scalable} V2X \gls{ndt}, capable of replicating the coexistence of heterogeneous technologies with unprecedented accuracy. By integrating the open source ray tracing engine Sionna RT into the full-stack ms-van3t simulator, we enable high-fidelity propagation modeling, scalable simulation across multiple communication technologies, and accurate interference tracking at the time-frequency level. Comparative simulations demonstrate that classical stochastic models produce oversimplified and often misleading results---e.g., up to 49\% mismatch in SINR-based reception decisions compared to the ray-based model in Van3Twin. The lack of consideration for the \gls{nlosv} scenarios in ms-van3t further broadens this mismatch, leading to 60\% disagreement in cases prone to vehicular blockage. Validation against real-world measurements confirms the improved propagation and application level fidelity of VaN3Twin. The use of ray tracing reduces the Packet Reception \gls{dr} by more than 50\% in rural scenarios and over 70\% in complex urban environments. These improvements position VaN3Twin as a valuable asset for site-specific validations, large-scale coexistence testing, and standardization support \firstrev{including, for example, the safe pre-deployment evaluation of new features introduced in 3GPP Rel. 18~\cite{rel18} such as ML-based scheduling, enhanced NR-V2X sidelink, and improved positioning, alongside AI model training with VaN3Twin-generated datasets and support in network planning operations}. Future work will offload additional PHY functions---e.g., waveform generation and multi-antenna systems—to GPUs, improving simulation speed and potentially accuracy by reducing the reliance on abstractions. 
\firstrev{Additionally, it will focus on systematically analyzing the trade-off between computational performance and simulation accuracy for example by tuning ray tracing parameters such as interaction types and maximum interaction depth, and by adapting the \gls{ndt} update rate to the dynamics of different environments.}
\secondrev{Another a viable direction for future work is assessing whether Universal Software Radio Peripheral (USRP) hardware and related software (such as OpenAirInterface or srsRAN) can be leveraged for the aim of performing additional field tests with \mbox{C-V2X} to further compare the outputs of VaN3Twin with real measured data. These tests will also be performed with dedicated C-V2X chipsets as compatible hardware will reach the market.} 
\secondrev{Finally, future work can also involve a deeper investigation on the real-time capabilities of VaN3Twin, also leveraging its emulation mode to interface with real-world connected vehicles.}

\section*{Acknowledgments}
This publication is part of the project PNRR-NGEU which has received funding from the MUR - DM 117/2023. This article was supported by the European Union under the Italian National Recovery and Resilience Plan (NRRP) of NextGenerationEU, partnership on “Telecommunications of the Future” (PE00000001 - program “RESTART”, Structural Project 6GWINET). This study was also carried out within the MOST – Sustainable Mobility National Research Center and received funding from the European Union Next-GenerationEU (PIANO NAZIONALE DI RIPRESA E RESILIENZA (PNRR) – MISSIONE 4 COMPONENTE 2, INVESTIMENTO 1.4 – D.D. 1033 17/06/2022, CN00000023).

\bibliographystyle{IEEEtran}
\bibliography{biblio.bib}

\vfill

\end{document}